\documentclass[a4paper,11pt]{article}
\usepackage{jcappub}
\usepackage{mathrsfs,amsmath,latexsym,amssymb,amsfonts,enumerate,graphicx,txfonts,subfigure,graphics}
\usepackage{xcolor}
\usepackage{etoolbox,orcidlink}
\allowdisplaybreaks
\newcommand{\GZU}{School of Physics, Guizhou University, Guiyang 550025, China}
\newcommand{\dd}{\mathrm{d}}

\title{Revisiting particle circular orbits as probes of black hole phase transitions}

\author[a]{Lei You\orcidlink{0009-0004-5900-3318},}
\author[a,1]{Jinsong Yang\orcidlink{0000-0003-4051-2767}\note{Corresponding author.}}

\affiliation[a]{\GZU}

\emailAdd{gs.lyou25@gzu.edu.cn}
\emailAdd{jsyang@gzu.edu.cn}

\abstract{Previous studies suggested that the particle circular orbit can serve as a probe of black hole phase transitions. However, these studies only identified this phenomenon by substituting the horizon radius $r_h$ with the circular orbit radius $r_c$ in thermodynamic state relations. Such simplistic substitution fails to uncover the underlying connection between black hole phase transitions and particle circular orbits. In this work, we successfully establish this profound intrinsic link by deriving a differential relation for $r_c$ that relates to thermodynamic parameters and the first law of black hole thermodynamics. Using this relation, we demonstrate that during a first-order phase transition, if $r_h$ experiences a discontinuous jump (such as in the small/large black hole phase transition), then $r_c$ must simultaneously undergo a discontinuous jump. This finding confirms that particle circular orbits can indeed serve as probes of first-order phase transitions. More importantly, we show that this phenomenon is a direct consequence of the nonzero latent heat inherent to first-order phase transitions. Finally, we demonstrate that the jump sizes $\Delta r_c$ and $\Delta r_h$ across the phase transition share the same critical exponent at the thermodynamic critical point, indicating that $\Delta r_c$ can serve as an order parameter for black hole phase transitions. Notably, this conclusion follows directly from the first law.}


\arxivnumber{2512.17642}

\begin{document}
\maketitle
\flushbottom

\section{Introduction}
\label{sec1}

Black hole thermodynamics provides a unique arena for exploring the interplay among gravity, quantum physics, and statistical mechanics. It began as a formal analogy between the laws of black hole mechanics and ordinary thermodynamics, and acquired genuine physical content through the generalized second law and Hawking radiation~\cite{Bekenstein:1973ur,Bardeen:1973gs,Bekenstein:1974ax,Hawking:1975vcx}. Subsequent developments connected it to broader areas of physics, for example microscopic state counting in string theory and holography in anti-de Sitter/conformal field theory (AdS/CFT)~\cite{Strominger:1996sh,Maldacena:1997re,Witten:1998qj}. Particularly influential progress has come from AdS black holes in the extended phase space, where the cosmological constant is interpreted as a pressure and the mass as an enthalpy~\cite{Kastor:2009wy,Dolan:2010ha,Cvetic:2010jb}. Within this framework, AdS black holes were found to possess a rich phase structure, including Van der Waals-like phase transitions and criticality, as well as more exotic behaviors in modified gravity and nonlinear electrodynamics~\cite{Banerjee:2011au,Banerjee:2011raa,Kubiznak:2012wp,Wei:2012ui,Cai:2013qga,Altamirano:2013uqa,Altamirano:2013ane,Frassino:2014pha,Wei:2014hba,Kubiznak:2016qmn}.

For a long time, possible observational signatures of these phase transitions have been actively sought. Early studies were devoted to exploring the connection between phase transitions and quasinormal modes (QNMs). For example, in Ref.~\cite{Liu:2014gvf} the authors found that when a charged AdS black hole undergoes the Van der Waals--like phase transition, the slope of the quasinormal frequency changes significantly. This indicates that QNMs can serve as a dynamical probe of thermodynamic phase transitions. This phenomenon was later confirmed by more works~\cite{Zou:2014sja,Chabab:2017knz,Zou:2017juz,Li:2017kkj}. In recent years, many works have noticed a connection between phase transitions and particle circular orbits. This makes it possible to observe phase transitions from black hole images. In Ref.~\cite{Wei:2017mwc}, the authors replaced the entropy $S$ (equivalent to the horizon radius $r_h$) in the temperature-entropy relation $T(S)$ of a charged AdS black hole with the photon circular orbit radius, i.e., the photon sphere radius $r_{\rm ps}$. They found that the resulting function $T(r_{\rm ps})$ still exhibits a Van der Waals-like phase transition curve. This indicates that during the phase transition, not only does the horizon radius $r_h$ jump from the small black hole branch to the large black hole branch, but $r_{\rm ps}$ also jumps from a smaller value to a larger one. Moreover, at the thermodynamic critical point, they found that the jump sizes $\Delta r_{\rm ps}$ has the universal critical exponent $1/2$. The authors later extended this idea to rotating Kerr--AdS black holes~\cite{Wei:2018aqm}. These results first demonstrated the possibility of probing black hole phase transitions via photon circular orbits. This variable-substitution approach was soon applied to other black hole spacetimes, further confirming the link between phase transitions and photon circular orbits~\cite{Zhang:2019tzi,NaveenaKumara:2019nnt,Xu:2019yub,Chabab:2019kfs,Du:2022quq,Ladino:2024ned,Ladino:2025oeq,Yang:2025xck}. However, these works rely only on a simple variable substitution, replacing the volume or entropy (equivalent to the horizon radius $r_h$) with the photon sphere radius $r_{\rm ps}$ in thermodynamic state relations. They therefore do not reveal a deeper connection between phase transitions and circular orbits. It is worth mentioning that Ref.~\cite{Zhang:2019tzi} gave a mathematical reason for the link between phase transitions and circular orbits. They proved that, in spherically symmetric spacetimes, the relation $r_{\rm ps}(r_h)$ must be monotonically increasing. As a result, the phase transition curve is preserved when $r_h$ is replaced by $r_{\rm ps}$ in thermodynamic state relations. This leads to the result that $r_{\rm ps}$ always jumps together with $r_h$ during the phase transition. This conclusion also holds for the innermost stable circular orbit (ISCO) radius $r_{\rm isco}$ of massive particles. We briefly review the mathematics in Appendix~\ref{appa}. The authors later gave a similar discussion for black hole shadows, which directly confirms that phase transitions can leave imprints on black hole images~\cite{Zhang:2019glo}. However, we note that their proof lacks rigor, as it does not account for the potential nonmonotonicity of $r_c(r_h)$ [including $r_{\rm ps}(r_h)$ and $r_{\rm isco}(r_h)$]. In such case, the link between phase transitions and circular orbits warrants further verification. Additionally, their proof is primarily mathematical and does not reveal the underlying physical mechanism.

Motivated by the above progress, this paper aims to further verify the universality of the connection between phase transitions and particle circular orbits, especially in the case where $r_c(r_h)$ exhibits nonmonotonic behavior. More importantly, we aim to reveal the underlying physical mechanism behind this connection. The paper is organized as follows. In section~\ref{sec2}, we review the formulas for particle circular orbits and use them to derive a monotonicity criterion for $r_c(r_h)$, namely $\dd r_c/\dd r_h$. In section~\ref{sec22}, we first use this criterion to analyze when $r_c(r_h)$ is monotonic and when it is nonmonotonic. We then prove whether $r_c$ always jumps with $r_h$, to verify the universality of the relation between phase transitions and circular orbits. In section~\ref{sec3}, taking the $d$-dimensional RN--AdS black hole as an example, we verify that $r_c(r_h)$ indeed exhibits both monotonic and nonmonotonic behaviors and that $r_c$ necessarily jumps at the phase transition. In section~\ref{sec4}, we prove that the order parameters defined by $r_c$ and $r_h$ have the same critical exponent. Finally, in section~\ref{sec5} we summarize and discuss our results. Throughout this work we employ natural units with $c=G=1$.

\section{Monotonicity criterion}
\label{sec2}

In this section, we first review the equations for circular orbits of massless and massive particles. We then use them to derive a monotonicity criterion for $r_c(r_h)$, namely $\dd r_c/\dd r_h$.

\subsection{Review of circular orbits}

Consider a static, spherically symmetric spacetime with the line element
\begin{equation}
 \dd s^2=-f(r;\lambda)\,\dd t^2+\frac{\dd r^2}{g(r;\lambda)}+r^2 \left(\dd \theta^2 + \sin^2\!\theta \, \dd \phi^2 \right),
 \label{eq:ybdg}
\end{equation}
where $f(r;\lambda)$ and $g(r;\lambda)$ are two smooth functions of the radial coordinate $r$, while $\lambda=(\lambda_1,\ldots,\lambda_n)$ denotes the set of black hole parameters, such as mass $M$, charge $q$, and so on. The Lagrangian for geodesic motion of test particles reads
\begin{equation}
 \mathcal{L}
 =\tfrac12\,g_{\mu\nu}\dot x^\mu\dot x^\nu
 =-\tfrac12\,\epsilon,\qquad
 \epsilon=
 \begin{cases}
 0, & \text{massless particle (photon)},\\
 1, & \text{massive particle}.
 \end{cases}
 \label{eq:l}
\end{equation}
Where the overdot denotes differentiation with respect to the affine parameter. Without loss of generality, we set $\theta=\pi/2$, restricting the motion to the equatorial plane. Then, the Lagrangian reduces to
\begin{equation}
 \mathcal{L}
 =\frac12\!\left[-f(r;\lambda)\,\dot t^{\,2}
 +\frac{\dot r^{\,2}}{g(r;\lambda)}
 +r^2\dot\phi^{\,2}\right]=-\frac12\,\epsilon.
 \label{eq:lcd}
\end{equation}
Since the metric is stationary and axisymmetric, the corresponding cyclic coordinates $t$ and $\phi$ give rise to two conserved quantities,
\begin{equation}
 E := -\frac{\partial\mathcal{L}}{\partial \dot t}
 = f(r;\lambda)\,\dot t,
 \qquad
 L := \frac{\partial\mathcal{L}}{\partial \dot\phi}
 = r^2 \dot\phi,
 \label{eq:el}
\end{equation}
which represent the particle's energy and angular momentum, respectively. Substituting eq.~\eqref{eq:el} into eq.~\eqref{eq:lcd} yields the following radial equation of motion
\begin{equation}
 \dot r^{\,2}
 =\frac{g(r;\lambda)}{f(r;\lambda)}E^2-g(r;\lambda)\!\left(\epsilon+\frac{L^2}{r^2}\right)
 \equiv\mathcal{R}(r).
 \label{eq:rd}
\end{equation}
The particle's circular orbits at $r=r_c$ are determined by
\begin{equation}
 \mathcal{R}(r_c)=0, \qquad
 \mathcal{R}'(r_c)=0,
 \label{eq:ygdtj}
\end{equation}
where the prime denotes differentiation with respect to $r$. Solving eq.~\eqref{eq:ygdtj} for a massless particle with $\epsilon=0$ gives
\begin{equation}
 \Phi_0(r;\lambda)
 :=r\,\partial_r f(r;\lambda)-2f(r;\lambda)=0,
 \qquad\text{at}\,r=r_{\rm ps}.
 \label{eq:phi0}
\end{equation}
The condition for an unstable photon sphere further requires
\begin{equation}
 \mathcal{R}''(r_{\rm ps})>0,
\end{equation}
which is equivalent to
\begin{equation}
 \Gamma_0(\lambda)
 :=\left.\partial_r\Phi_0(r;\lambda)\right|_{r_{\rm ps}}<0.
 \label{eq:g0}
\end{equation}
The ISCO of a massive particle with $\epsilon=1$ marks the transition between stable and unstable circular motion and is defined by the marginal stability condition
\begin{equation}
 \mathcal{R}(r_{\rm isco})
 =\mathcal{R}'(r_{\rm isco})
 =\mathcal{R}''(r_{\rm isco})=0,
\end{equation}
which can be recast into a universal form,
\begin{align}
 \Phi_1(r;\lambda)
 :=& r\,g(r;\lambda)\,f(r;\lambda)\,\partial_r^2 f(r;\lambda)
 - 2\,r\,\partial_r g(r;\lambda)\,f(r;\lambda)\,\partial_r f(r;\lambda)
 + 3\,g(r;\lambda)\,f(r;\lambda)\,\partial_r f(r;\lambda)\notag\\
 &+2\,r\,\partial_r g(r;\lambda)\,f(r;\lambda)
 -2\,r\,g(r;\lambda)\,\partial_r f(r;\lambda)=0,\qquad\text{at}\,r=r_{\rm isco}.
 \label{eq:phi1}
\end{align}
To ensure that circular orbits are stable outside ISCO but unstable inside, one further requires its radius $r_{\rm isco}$ to satisfy
\begin{equation}
 \Gamma_1(\lambda)
 :=\left.\partial_r\Phi_1(r;\lambda)\right|_{r_{\rm isco}}>0.
 \label{eq:g1}
\end{equation}
The detailed derivations of eqs.~\eqref{eq:phi0}, \eqref{eq:g0}, \eqref{eq:phi1}, and \eqref{eq:g1} are provided in Appendix~\ref{appg}. Summarizing the above conditions, we have
\begin{equation}
 \Phi_\epsilon(r;\lambda)=0,\qquad\text{at}\,
 r=r_c(\epsilon):=
 \begin{cases}
 r_{\rm ps}, & \epsilon=0,\\
 r_{\rm isco}, & \epsilon=1,
 \end{cases}
\end{equation}
and the corresponding stability criteria
\begin{equation}
 \Gamma_\epsilon\,
 \begin{cases}
 <0, & \epsilon=0,\\[4pt]
 >0, & \epsilon=1.
 \end{cases}
 \label{eq:g0g1}
\end{equation}

\subsection{Derivation of the monotonicity criterion}

To investigate the monotonicity properties of the relation between the circular orbit radius $r_c$ and the event horizon radius $r_h$, we first express them in terms of the previously defined functions $\Phi_\epsilon(r;\lambda)$ and $f(r;\lambda)$
\begin{equation}
 \Phi_\epsilon(r_{c};\lambda)=0,
 \qquad
 f(r_h;\lambda)=0.
 \label{eq:ksfc}
\end{equation}
According to the implicit function theorem, if
\begin{equation}
 \left.\partial_r \Phi_\epsilon\right|_{(r_{\mathrm{c}};\lambda)} \neq 0,
 \qquad
 \left.\partial_r f\right|_{(r_h;\lambda)} \neq 0,
 \label{eq:yhsyq}
\end{equation}
then there exist locally smooth functions $r_{\mathrm{c}}(\lambda)$ and $r_h(\lambda)$ in the neighborhoods of $(r_{\mathrm{c}};\lambda)$ and $(r_h;\lambda)$, respectively. Fortunately, according to eqs.~\eqref{eq:g0} and \eqref{eq:g1}, one finds
\begin{equation}
 \left.\partial_r \Phi_\epsilon\right|_{(r_{\mathrm{c}};\lambda)}=\Gamma_\epsilon\neq 0,
 \label{eq:gneqz}
\end{equation}
and
\begin{equation}
 \left.\partial_r f\right|_{(r_h;\lambda)}\propto 2\kappa\, \neq\, 0,
\end{equation}
where
\begin{equation}
 \kappa=\frac12\lim_{r\to r_h}\sqrt{\frac{g(r;\lambda)}{f(r;\lambda)}}\,\partial_r f(r;\lambda)
\end{equation}
is the surface gravity of the black hole, which is positive for any nonextremal case. Hence, the conditions~\eqref{eq:yhsyq} are automatically satisfied.

Based on this, eq.~\eqref{eq:ksfc} can in principle be differentiated with respect to $\lambda$. Since the parameter set $\lambda$ generally contains several components, it is convenient to introduce a smooth path $\lambda(\tau)$ in the parameter space, along which all quantities become functions of a single real variable $\tau$. This allows us to directly apply the chain rule, yielding
\begin{equation}
 \frac{\dd}{\dd \tau}\,\Phi_\epsilon\big(r_{\mathrm{c}}(\tau);\lambda(\tau)\big)
 = \partial_{r_c}\Phi_\epsilon\,\dot {r_c}
 + \langle \partial_\lambda \Phi_\epsilon,\,\dot\lambda \rangle = 0,
 \label{eq:wf1}
\end{equation}
\begin{equation}
 \frac{\dd}{\dd \tau}\,f\big(r_h(\tau);\lambda(\tau)\big)
 = \partial_{r_h}f\,\dot r_h
 + \langle \partial_\lambda f,\,\dot\lambda \rangle = 0,
 \label{eq:wf2}
\end{equation}
where the overdot denotes differentiation with respect to $\tau$, and $\langle\partial_\lambda F,\dot{\lambda}\rangle\equiv\sum_{i=1}^n (\partial_{\lambda_{i}} F)\,\dot{\lambda}_i$. Combining eqs.~\eqref{eq:wf1} and \eqref{eq:wf2}, we finally obtain
\begin{equation}
 \frac{\dd r_c}{\dd r_h}
 = \frac{\dot r_{\mathrm{c}}}{\dot r_h}
 = \frac{2 \hat{\kappa}}{\Gamma_\epsilon}\,
 \frac{\langle \partial_\lambda \Phi_\epsilon,\,\dot{\lambda}\rangle}
 {\langle \partial_\lambda f,\,\dot{\lambda}\rangle},
 \qquad (\dot r_h\neq 0),
 \label{eq:pj}
\end{equation}
where $\hat{\kappa}\equiv\kappa \lim_{r\to r_h}\sqrt{{f(r;\lambda)}/{g(r;\lambda)}}$. When $f(r;\lambda)=g(r;\lambda)$, one has $\hat{\kappa}=\kappa=1/2\, \partial_r f(r_h;\lambda)$, recovering the familiar expression. Moreover, we have written $\Phi_\epsilon$ and $f$ as shorthand for $\Phi_\epsilon(r_{\mathrm{c}};\lambda)$ and $f(r_h;\lambda)$, respectively, and this convention will be used throughout the paper.

The criterion~\eqref{eq:pj} shows that the monotonicity of $r_{\mathrm{c}}(r_h)$ is governed by two independent factors. The first factor, $2\hat{\kappa}/\Gamma_\epsilon$, is purely geometric, determined by the intrinsic properties of the black hole and its circular orbits. As discussed earlier, it remains nonvanishing and therefore does not affect the monotonic behavior of $r_{\mathrm{c}}(r_h)$. The second factor, $\langle\partial_\lambda\Phi_\epsilon,\dot{\lambda}\rangle / \langle\partial_\lambda f,\dot{\lambda}\rangle$, encodes how the parameters evolve along a chosen path in the thermodynamic space. It therefore determines whether the monotonic behavior of $r_{\mathrm{c}}(r_h)$ is preserved or violated.

\section{Monotonicity and jumps of $r_c(r_h)$ across phase transitions}
\label{sec22}

In this section we further analyze the criterion~\eqref{eq:pj} under appropriate simplifying assumptions. The analysis has two parts. We first examine whether $r_c(r_h)$ can become nonmonotonic. If such nonmonotonicity arises, we then examine whether $r_c$ still undergoes a discontinuous jump together with $r_h$ at the phase transition.

\subsection{Monotonicity analysis of $r_c(r_h)$}

For the criterion~\eqref{eq:pj}, a simple choice of the path parameter is the horizon radius $r_h$, which ensures $\dot r_h=1\neq 0$. With this choice, the criterion can be reduced to
\begin{equation}
 \frac{\dd r_c}{\dd r_h}
 =-\frac{1}{\Gamma_\epsilon}\,\left(\partial_M \Phi_\epsilon\,\frac{\dd M}{\dd r_h}+\partial_P \Phi_\epsilon\,\frac{\dd P}{\dd r_h}+\sum_j \partial_{\lambda_j} \Phi_\epsilon\,\frac{\dd \lambda_j}{\dd r_h}\right),
 \label{eq:pjj1}
\end{equation}
where $M$ denotes the black hole mass, $P=-\Lambda/(8\pi)$ is the pressure associated with the negative cosmological constant $\Lambda$ in the extended phase space, and $\{\lambda_j\}$ collects all other metric parameters except $M$ and $P$. To keep the discussion as model independent as possible while remaining analytically tractable, we work on a fixed-$\{\lambda_j\}$ slice of the parameter space, i.e., $\dd\lambda_j=0$. This setting covers a wide range of well-known phase transitions, including Van der Waals-like small/large black hole phase transitions, reentrant phase transitions, and triple-point behavior. If additional parameters are allowed to vary, the phase diagram becomes higher dimensional and may exhibit new multicritical structures. Such a more general but also more complex situation will not be considered here. Under this simplification, the criterion~\eqref{eq:pjj1} further reduces to
\begin{equation}
 \frac{\dd r_c}{\dd r_h}
 =-\frac{1}{\Gamma_\epsilon}\left(
 \partial_M \Phi_\epsilon\,\frac{\dd M}{\dd r_h}
 +\partial_P \Phi_\epsilon\,\frac{\dd P}{\dd r_h}\right).
 \label{eq:zy1}
\end{equation}
Before presenting our analysis, we make the following assumptions. Since the mass $M$ is the primary source of gravity that fixes the background geometry and hence affects the circular orbit radius, we assume $\partial_M \Phi_\epsilon \neq 0$. For $\partial_P \Phi_\epsilon$, we likewise assume $\partial_P \Phi_\epsilon \neq 0$ whenever the circular orbit radius depends on the pressure $P$. However, it should be noted that $P$ may have no effect on the circular orbit radius. For example, if $P$ enters the metric only through a term of the form $P r^{2}$, then $\Phi_0$ can be independent of $P$. A familiar case is the Sch--AdS black hole, for which the photon sphere radius $r_{\rm ps}=3M$ is independent of $P$. Thus, in such situations, $\partial_P \Phi_\epsilon$ vanishes identically, i.e., $\partial_P \Phi_\epsilon \equiv 0$.

In Ref.~\cite{Zhang:2019tzi}, without imposing any thermodynamic path constraint, the authors claimed that $r_c(r_h)$ must be monotonic, but the argument is not fully rigorous. Below we use the criterion~\eqref{eq:zy1} to clarify the assumptions that guarantee a monotonic $r_c(r_h)$. From the criterion~\eqref{eq:zy1}, it is clear that when both $M$ and $P$ vary, the two terms in parentheses can compete and may drive $\dd r_c/\dd r_h$ to cross zero. By contrast, if only $M$ or only $P$ is allowed to vary, the situation simplifies: whether $\dd r_c/\dd r_h$ crosses zero is controlled solely by whether $\dd M/\dd r_h$ or $\dd P/\dd r_h$ crosses zero. Interestingly, we find that the zero-crossing behaviors of $\dd M/\dd r_h$ and $\dd P/\dd r_h$ are closely tied to the first law of black hole thermodynamics. In the extended phase space, the mass $M$ is interpreted as the enthalpy, and the first law takes the form
\begin{equation}
 \dd M
 = T\,\dd S
 + V\,\dd P
 + \sum_{k} X_k\,\dd Y_k \,,
\end{equation}
where $S$ is the black hole entropy, $T$ is the thermodynamic temperature conjugate to $S$, and $V$ is the thermodynamic volume conjugate to $P$. In addition, $\{Y_k\}$ denote other thermodynamic variables associated with parameters in the black hole metric, and $\{X_k\}$ are the thermodynamic quantities conjugate to $\{Y_k\}$. Since we fix all black hole parameters except $M$ and $P$, it follows that $\dd Y_k=0$. Therefore, the first law reduces to
\begin{equation}
 \dd M = T\,\dd S + V\,\dd P.
\end{equation}
Dividing both sides by $\dd r_h$ then gives
\begin{equation}
 \frac{\dd M}{\dd r_h}
 = T\,S'(r_h) + V\,\frac{\dd P}{\dd r_h}.
 \label{eq:zy2}
\end{equation}
It is worth emphasizing that the black hole temperature $T$ and volume $V$ must be positive. Moreover, $S'(r_h)$ is typically positive. For example, for $S=\pi r_h^{2}$ one has $S'(r_h)=2\pi r_h>0$.

From eq.~\eqref{eq:zy2}, it can be found that fixing the pressure ($\dd P=0$) enforces $\dd M/\dd r_h>0$. Conversely, fixing the mass ($\dd M=0$) enforces $\dd P/\dd r_h<0$. Substituting these sign constraints into the criterion~\eqref{eq:zy1} leads to a simple conclusion: along thermodynamic processes with either $M$ or $P$ held fixed, $\dd r_c/\dd r_h\neq 0$, and hence $r_c(r_h)$ is necessarily monotonic. In thermodynamics, the simplest processes satisfying these conditions are the isenthalpic and isobaric ones. Moreover, it is worth emphasizing that this conclusion is enforced by the first law, indicating an underlying link between spactime geometry and black hole thermodynamics. To our knowledge, this point has not been recognized in previous related studies.

Next, we further show that $r_c(r_h)$ can become nonmonotonic, which necessarily requires that both $M$ and $P$ vary. Under this setting, three cases can arise:
\begin{enumerate}[(i)]
\item when $\partial_P \Phi_\epsilon \equiv 0$ (often happens for $r_{\rm ps}$), whether $\dd {r_c}/ \dd {r_h}$ crosses zero is controlled by $\dd M / \dd {r_h}$. Equation~\eqref{eq:zy2} shows that $\dd M / \dd {r_h}$ can cross zero if $\dd P / \dd {r_h}$ is sufficiently negative. Otherwise, it cannot cross zero. This zero-crossing behavior determines whether $r_c(r_h)$ is monotonic or nonmonotonic. In the next section, we will show that both possibilities can occur.
\item when $\partial_P\Phi_\epsilon\neq 0$ and $\partial_P\Phi_\epsilon$ has the opposite sign to $\partial_M\Phi_\epsilon$, $\dd r_c/\dd{r_h}$ may cross zero if $\dd M/\dd{r_h}$ and $\dd P/\dd{r_h}$ have the same sign. Otherwise, it cannot cross zero. Interestingly, Equation~\eqref{eq:zy2} allows $\dd M/\dd{r_h}$ and $\dd P/\dd{r_h}$ to share the same sign. If $\dd M/\dd{r_h}<0$, then $\dd P/\dd{r_h}$ must also be negative. If $\dd P/\dd{r_h}>0$, then $\dd M/\dd{r_h}$ must be positive as well. This indicates that $r_c(r_h)$ can be nonmonotonic, and we will show in the next section that this possibility indeed occurs.
\item when $\partial_P\Phi_\epsilon\neq 0$ and $\partial_P\Phi_\epsilon$ has the same sign as $\partial_M\Phi_\epsilon$, a zero crossing of $\dd {r_c}/ \dd {r_h}$ is unavoidable if both $\dd M/\dd{r_h}$ and $\dd P/\dd{r_h}$ have zeros. According to eq.~\eqref{eq:zy2}, at a zero of $\dd M/\dd{r_h}$, say $r_{h,m}$, one has $\dd P/\dd{r_h}<0$, while at a zero of $\dd P/\dd{r_h}$, say $r_{h,p}$, one has $\dd M/\dd{r_h}>0$. Substituting these signs into the criterion~\eqref{eq:zy1}, one immediately finds that $\dd {r_c}/ \dd {r_h}$ has opposite signs at $r_{h,m}$ and $r_{h,p}$. Therefore, $\dd {r_c}/ \dd {r_h}$ must cross zero between $r_{h,m}$ and $r_{h,p}$, implying that $r_c(r_h)$ is nonmonotonic.
\end{enumerate}
It should be pointed out that, since $M$ and $P$ typically affect the circular orbit radius in opposite ways, the third case is unlikely to occur. In any event, these three cases indicate that the first law fully allows $r_c(r_h)$ to be nonmonotonic. As will be shown in the next section, nonmonotonic $r_c(r_h)$ can indeed arise.
\begin{figure}[htbp]
 \centering
 \includegraphics[width=0.6\linewidth]{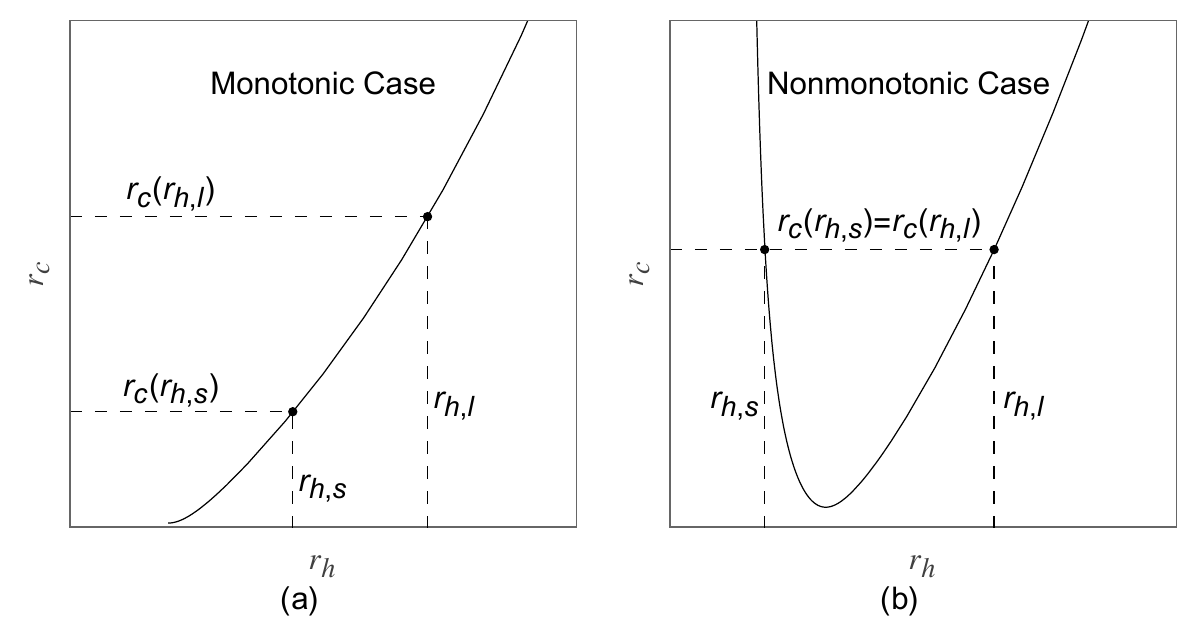}

 \caption{Panels (a) and (b) illustrate the monotonic and nonmonotonic cases of $r_c(r_h)$, respectively. The symbols $r_{h,s}$ and $r_{h,l}$ denote the horizon radii before and after the phase transition, respectively.}
 \label{mnm}
\end{figure}

\subsection{Jump of $r_c$ at the phase transition}

Having confirmed that $r_c(r_h)$ may be nonmonotonic, we now examine whether $r_c$ must jump together with $r_h$ at a phase transition. If $r_c(r_h)$ is monotonic, then $r_c(r_{h,s})\neq r_c(r_{h,l})$, so $r_c$ necessarily jumps when $r_h$ jumps [see figure~\ref{mnm}(a)]. Here $r_{h,s}$ and $r_{h,l}$ denote the horizon radii before and after the phase transition, respectively. By contrast, if $r_c(r_h)$ is nonmonotonic, it is in principle possible that $r_c(r_{h,s})=r_c(r_{h,l})$, in which case $r_c$ does not jump even though $r_h$ does [see figure~\ref{mnm}(b)]. Next, we will show that, in all cases, $r_c$ must jump together with $r_h$ at the phase transition.

\begin{figure}[htbp]
 \centering
 \includegraphics[width=0.3\linewidth]{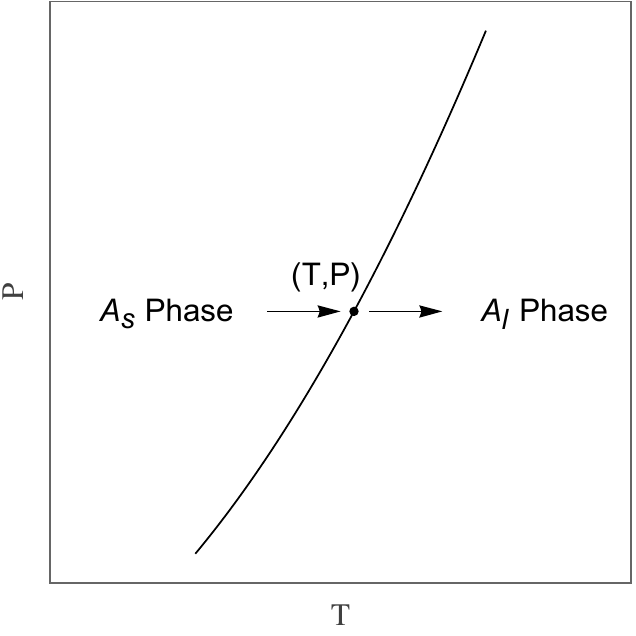}

 \caption{Schematic of a coexistence line in the $P$-$T$ plane. The $A_s$ and $A_l$ phases coexist on the black solid line.}
 \label{xt}
\end{figure}

In thermodynamics, phase transitions can be identified by nonanalytic behavior of a suitable thermodynamic potential. A standard classification is the Ehrenfest scheme, which defines the order of a phase transition by the lowest-order derivative of the potential that becomes discontinuous at the transition point. In the extended phase space, phase transitions that cause a discontinuous jump in $r_h$ are typically first order. For example, the small/large black hole phase transition in the RN--AdS spacetime is described by the Gibbs free energy $G(T,P)$. It exhibits the characteristic swallowtail structure in the $G$-$T$ or $G$-$P$ plane~\cite{Kubiznak:2012wp}. At the phase transition point, $G$ is continuous while its first derivative is discontinuous, so the phase transition is first order. However, it should be noted that recent studies have also found that a jump in $r_h$ can occur in a zeroth-order phase transition~\cite{Dehyadegari:2017flm,Stetsko:2018jqt}. For instance, Ref.~\cite{Dehyadegari:2017flm} studied the thermodynamics of charged dilaton black holes in the extended phase space and found that, at the small/large black hole phase transition, the dilaton-electromagnetic coupling constant leads to a finite jump of the Gibbs free energy. This indicates that the phase transition is zeroth order. To cover both possibilities, we consider zeroth and first order phase transitions. These two types of phase transitions are usually represented in the $P$-$T$ phase diagram, as shown in figure~\ref{xt}. The black solid curve is the coexistence line, which separates two distinct phases of the system. When the system undergoes a transition from phase $A_s$ to phase $A_l$, the ordered pair $(P,T)$ formed by the system's pressure and temperature lies on the coexistence line. In other words, the two phases share the same pressure and temperature at the phase transition. For a first order phase transition, they also have the same Gibbs free energy.

Our analysis starts from the Gibbs free energy $G(T,P)$, which in the extended phase space takes the form
\begin{equation}
 G=M-TS.
\end{equation}
Across the phase transition, one has
\begin{equation}
 G_l-G_s=M_l-M_s-\left(T_l\, S_l - T_s\, S_s\right),
\end{equation}
where the subscripts $s$ and $l$ denote the values of the physical quantities before and after the phase transition, respectively. Rearranging the above relation and noting that the temperature is unchanged across the phase transition, one obtains
\begin{equation}
 M_l-M_s=G_l-G_s+T_p\left(S_l-S_s\right),
 \label{eq:mgts}
\end{equation}
where $T_p\equiv T_s=T_l$ is the phase transition temperature. For a first order phase transition, the above relation reduces to a simple form,
\begin{equation}
 M_l-M_s=T_p\left(S_l-S_s\right),
 \label{eq:ddm}
\end{equation}
where the right-hand side is precisely the latent heat, namely the heat exchanged during the phase transition while the temperature stays fixed. For a first order transition it is nonzero. The reason is simple: in the $G$-$T$ plane, the first derivative of the Gibbs free energy is the entropy,
\begin{equation}
 \frac{\partial G}{\partial T}=-S.
\end{equation}
A first-order phase transition is defined by a discontinuity in this first derivative. Therefore the entropy must jump across the transition, i.e., $S_s\neq S_l$. Meanwhile, noting that the black hole temperature must be positive ($T_p>0$), one obtains $M_l-M_s\neq 0$. Therefore, in a first order phase transition the black hole mass must change, and the change is exactly the latent heat. It should be pointed out that, if the system has a second-order phase transition, then at the critical point the distinction between the two phases disappears and the first order phase transition no longer occurs. This causes the latent heat to vanish, and hence $M_l-M_s=0$ at the critical point. For a zeroth-order phase transition, eq.~\eqref{eq:mgts} can no longer be simplified to a compact form. Nevertheless, one can still obtain a generic conclusion under mild assumptions: If $G_l-G_s$ and $S_l-S_s$ do not vanish simultaneously and share the same sign, then $M_l-M_s\neq 0$ must hold. The zeroth-order phase transition reported in Ref.~\cite{Dehyadegari:2017flm,Stetsko:2018jqt} satisfies this condition.

Keeping in mind the above conclusions for $M_s$ and $M_l$, we now return to the criterion~\eqref{eq:zy1}. By multiplying both sides by $\dd r_h$, we obtain its differential form,
\begin{equation}
 \dd r_c
 =-\frac{1}{\Gamma_\epsilon}\left(
 \partial_M \Phi_\epsilon\,\dd M
 +\partial_P \Phi_\epsilon\,\dd P\right).
\end{equation}
Since eq.~\eqref{eq:gneqz} guarantees that $r_c=r_c\left(M,P,\{\lambda_j\}\right)$ is locally smooth, the above relation is the total differential of $r_c$ on the fixed-$\{\lambda_j\}$ slice. Therefore, when the system crosses from $(M_s,P_s)$ to $(M_l,P_l)$ during a phase transition, the integral
\begin{equation}
 \Delta r_c\equiv r_{c,l}-r_{c,s}
 =\int_{\gamma}\dd r_c
 =-\int_{\gamma}\frac{1}{\Gamma_\epsilon}\Bigl(
 \partial_M\Phi_\epsilon\,\dd M+\partial_P\Phi_\epsilon\,\dd P
 \Bigr)
 \label{eq:drcpath}
\end{equation}
yields the change in $r_c$, where $\gamma$ is any path in the $(M,P)$ plane connecting $(M_s,P_s)$ and $(M_l,P_l)$. $r_{c,s}$ and $r_{c,l}$ denote $r_c(r_{h,s})$ and $r_c(r_{h,l})$, respectively. This quantity $\Delta r_c$ is exactly what we want to characterize. If $\Delta r_c\neq 0$, then $r_c$ must jump together with $r_h$ at the phase transition. Conveniently, during the phase transition the pressure stays fixed, which selects a natural path $\gamma$ with $P=P_s=P_l\equiv P_p$ and hence $\dd P=0$. Along this path eq.~\eqref{eq:drcpath} reduces to
\begin{equation}
 \Delta r_c
 =-\int_{M_s}^{M_l}\frac{1}{\Gamma_\epsilon}\,\partial_M\Phi_\epsilon\,\dd M,
 \label{eq:drc}
\end{equation}
where $1/\Gamma_\epsilon$ and $\partial_M\Phi_\epsilon$ should be understood as functions of $(M,P_p,\{\lambda_j\})$. Note that $\Gamma_\epsilon\neq 0$ is given in eq.~\eqref{eq:g0g1}, and we also assume $\partial_M\Phi_\epsilon\neq 0$. Therefore, the above integral expression shows that $\Delta r_c\neq 0$ as long as $M_s\neq M_l$. Moreover, it should be pointed out that this conclusion holds for both monotonic and nonmonotonic $r_c(r_h)$.

Finally, we summarize the above results as follows. For a first-order phase transition described by the Gibbs free energy, $r_c$ must jump together with $r_h$ at the phase transition, regardless of whether $r_c(r_h)$ is monotonic or nonmonotonic. This follows from the nonzero latent heat in a first-order phase transition. At the critical point, the first-order transition terminates, and neither $r_c$ nor $r_h$ exhibits a discontinuous jump. For a zeroth-order phase transition, $r_c$ must also jump with $r_h$ provided that $G_l-G_s$ and $S_l-S_s$ are not simultaneously zero and have the same sign. Moreover, once the signs of $1/\Gamma_\epsilon$ and $\partial_M\Phi_\epsilon$ and the ordering between $M_s$ and $M_l$ are known, eq.~\eqref{eq:drc} directly determines whether $r_{\rm c}$ jumps upward or downward. The above results, in more general cases, confirm the universality of the connection between black hole phase transitions and particle circular orbits, and reveal the deeper physical origin of this connection. In the next section, we take $d$-dimensional charged AdS black holes as examples to confirm that $r_c(r_h)$ can be either monotonic or nonmonotonic. We then verify that even when $r_c(r_h)$ is nonmonotonic, $r_c$ still jumps together with $r_h$ at the phase transition.

\section{Examples: $d$-dimensional charged AdS black holes}
\label{sec3}

In $d$-dimensional spacetime, a static and spherically symmetric charged AdS black hole is described by the line element
\begin{equation}
 \dd s^2 = -f(r)\,\dd t^2 + \frac{\dd r^2}{f(r)} + r^2\,\dd\Omega_{d-2}^2,
 \label{eq:rnads}
\end{equation}
where $\dd\Omega_{d-2}^2$ denotes the line element of a unit $(d-2)$-sphere, and the metric function $f(r)$ takes the general form
\begin{equation}
 f(r) = 1 - \frac{c_d M}{r^{d-3}} + \frac{q^2}{r^{2(d-3)}} + \frac{16\pi P r^2}{(d-1)(d-2)}\,,
\end{equation}
where $q$ is the charge parameter. The coefficients
\begin{equation}
 c_d=\frac{16\pi}{(d-2)\omega_{d-2}},
\end{equation}
depend only on the spacetime dimension, with $\omega_{d-2}$ being the area of the unit $(d-2)$-sphere
\begin{align}
 \omega_{d-2}=\frac{2\pi^{(d-1)/2}}{\Gamma[(d-1)/2]},
\end{align}
where $\Gamma(x)$ is the Gamma function. The metric~\eqref{eq:rnads} reproduces several well-established limits:
\begin{itemize}
\item Case $q=0$: The $d$-dimensional Sch--AdS black hole;
\item Case $q\neq0$: The RN--AdS black hole;
\item Case $P\to0$: The asymptotically flat Schwarzschild or RN solution in $d$ dimensions.
\end{itemize}

\begin{figure}[htbp]
 \centering
 \begin{minipage}{0.48\linewidth}
 \centering
 \includegraphics[width=\linewidth]{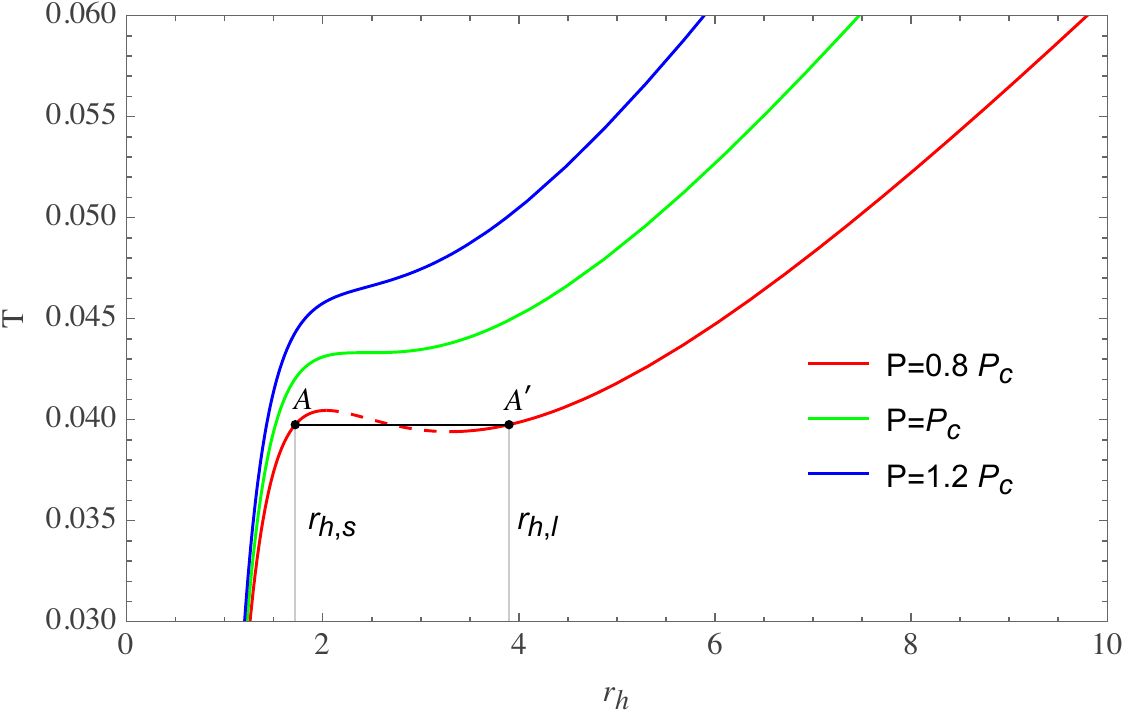}
 \end{minipage}
 \hfill
 \begin{minipage}{0.48\linewidth}
 \centering
 \includegraphics[width=\linewidth]{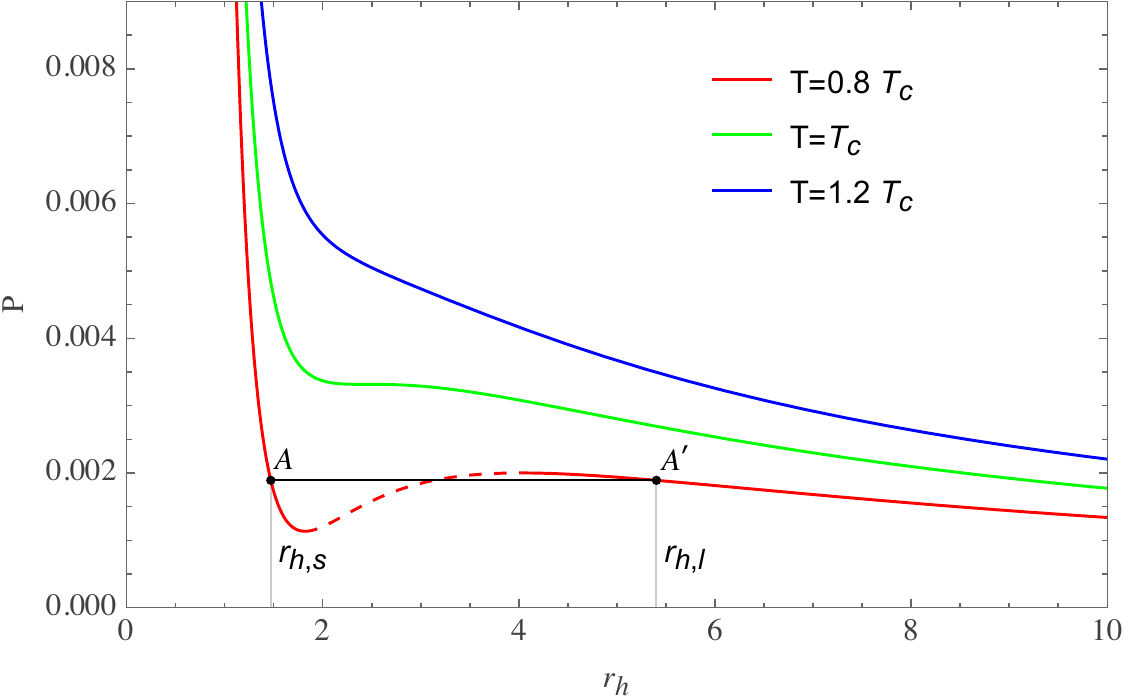}
 \end{minipage}

 \caption{Isobaric $T(r_h)$ curves (left panel) and isothermal $P(r_h)$ curves (right panel) of the RN--AdS black hole with $q=1$, shown for three different pressures and three different temperatures, respectively.}
 \label{prhtrh}
\end{figure}

\begin{figure}[htbp]
 \centering
 \includegraphics[width=0.48\linewidth]{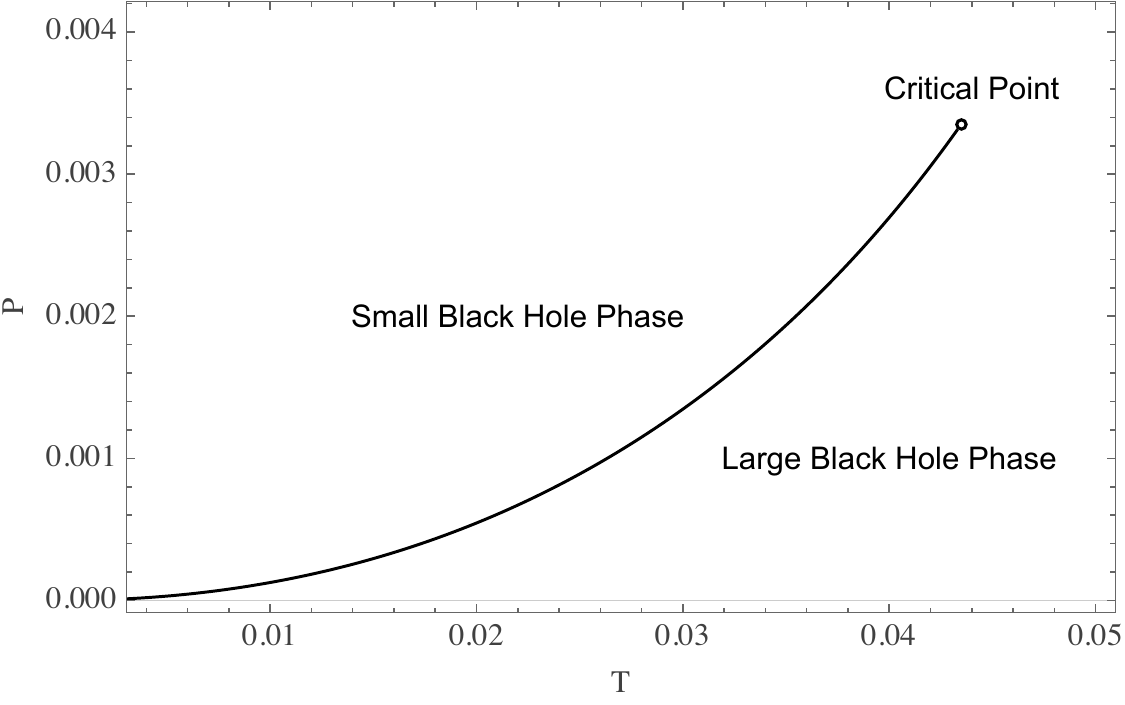}

 \caption{Coexistence line of the RN--AdS black hole in the $P$-$T$ plane for $q=1$. The critical point satisfies $P=P_c$ and $T=T_c$.}
 \label{pt}
\end{figure}

It is well known that $d$-dimensional charged AdS black holes exhibit a Van der Waals-like phase transition, which is a standard first-order phase transition~\cite{Kubiznak:2012wp,Gunasekaran:2012dq}. Here we briefly review this phenomenon in four dimensions. The black hole temperature is
\begin{equation}
 T(r_h,q,M,P)=\frac{\partial_r f(r_h)}{4 \pi}=\frac{-3q^{2}+3Mr_h+8\pi P\,r_h^{4}}{6\pi r_h^{3}}.
\end{equation}
The mass $M$ can be further determined from the horizon condition $f(r_h)=0$ as
\begin{equation}
 M(r_h,q,P)=\frac{3q^{2}+3r_h^{2}+8\pi P\,r_h^{4}}{6r_h}.
\end{equation}
Substituting this into the above expression yields the familiar form
\begin{equation}
 T(r_h,q,P)=\frac{-q^{2}+r_h^{2}+8\pi P\,r_h^{4}}{4\pi r_h^{3}}.
 \label{eq:trh}
\end{equation}
Solving the above relation for the pressure gives
\begin{equation}
 P(r_h,q,T)=\frac{q^{2}-r_h^{2}+4\pi T r_h^{3}}{8\pi r_h^{4}}.
 \label{eq:prh}
\end{equation}
In the canonical ensemble with fixed charge, eqs.~\eqref{eq:trh} and \eqref{eq:prh} describe the isobaric and isothermal curves, respectively, as displayed in figure~\ref{prhtrh}. In this figure, $P_c$ and $T_c$ denote the critical pressure and critical temperature, which are obtained by solving
\begin{equation}
 \frac{\partial P}{\partial r_h}
 =\frac{\partial^2 P}{\partial r_h^2}=0,
 \qquad\text{or}
 \qquad
 \frac{\partial T}{\partial r_h}
 =\frac{\partial^2 T}{\partial r_h^2}=0.
\end{equation}
For the RN--AdS black hole, the above equations yield
\begin{equation}
 r_{h,c} = \sqrt{6}\,q,\qquad
 P_{c} = \frac{1}{96\pi\,q^{2}},\qquad
 T_{c} = \frac{1}{3\sqrt{6}\,\pi\,q},
 \label{eq:rhpt}
\end{equation}
where $r_{h,c}$ is the horizon radius at the critical point. The figures clearly show that for $P<P_c$ and $T<T_c$, both the isobaric and isothermal curves exhibit the characteristic Van der Waals–type S shape. This indicates that the black hole will skip the unstable branch (red dashed line) and undergo a first-order phase transition directly from point A to point A$'$. During the phase transition, the temperature and pressure remain unchanged, while the horizon radius jumps from $r_{h,s}$ to $r_{h,l}$, which can be determined by the Maxwell equal-area construction. It is worth noting that, strictly speaking, the Maxwell equal-area law applies only to $T(S)$ and $P(V)$, namely,
\begin{equation}
 \int_{S_s}^{S_l} T(S)\,\dd S
 = T_p\,\left(S_l-S_s\right),
 \label{eq:mks1}
\end{equation}
and
\begin{equation}
 \int_{V_s}^{V_l} P(V)\,\dd V
 = P_p\,\left(V_l-V_s\right),
 \label{eq:mks2}
\end{equation}
where $S=\pi\,r_h^2$ is the entropy conjugate to $T$ and $V=4\pi\,r_h^3/3$ is the thermodynamic volume conjugate to $P$. $T_p$ and $P_p$ denote the phase transition temperature and pressure. When working instead with $T(r_h)$ and $P(r_h)$, the above relations become
\begin{equation}
 \int_{r_{h,s}}^{r_{h,l}} T(r_h)\,\frac{\dd S}{\dd r_h}\,\dd r_h
 = \pi\,T_p\left(r_{h,l}^2-r_{h,s}^2\right),
 \label{eq:mks3}
\end{equation}
and
\begin{equation}
 \int_{r_{h,s}}^{r_{h,l}} P(r_h)\,\frac{\dd V}{\dd r_h}\,\dd r_h
 = \frac{4\pi}{3}\,P_p\left(r_{h,l}^3-r_{h,s}^3\right).
 \label{eq:mks4}
\end{equation}
This implies that, on the $T(r_h)$ or $P(r_h)$ curves, the areas of the two closed regions enclosed by the line segment $AA'$ and the S-shaped curve are not equal, as illustrated in figure~\ref{prhtrh}. Moreover, figure~\ref{pt} plots the coexistence line in the $P$-$T$ plane, where each point gives the phase transition pressure and temperature. At the endpoint, the distinction between the two phases disappears and the first-order phase transition no longer occurs.

Next, we apply the criterion~\eqref{eq:zy1} to the spacetime defined by metric~\eqref{eq:rnads} and analyze, in detail, the monotonicity of $r_{\rm c}(r_h)$ along both isobaric and isothermal processes, as well as whether $r_{\rm c}$ jumps together with $r_h$ at the phase transition.

\subsection{Isobaric process}

Along the isobaric process, the charge $q$ and the pressure $P$ in the metric~\eqref{eq:rnads} are held fixed, and the thermodynamic evolution is driven by variations of the mass $M$. This corresponds to the fixed-$P$ case discussed in section~\ref{sec22}. In this case, the criterion~\eqref{eq:zy1} reduces to
\begin{equation}
 \frac{\dd r_{\mathrm{c}}}{\dd r_h}
 = -\,\frac{1}{\Gamma_\epsilon}\,
 \frac{\dd M}{\dd r_h}\,
 \partial_{M}\Phi_\epsilon.
 \label{eq:pj1}
\end{equation}
Note that the first law now enforces
\begin{equation}
 \frac{\dd M}{\dd r_h}
 = T\,S'(r_h)
 = 2\pi r_h\,T>0.
 \label{eq:dm1}
\end{equation}
For the metric~\eqref{eq:rnads}, we obtain
\begin{equation}
 \begin{aligned}
 \partial_M \Phi_0
 &= (d-1)\frac{c_d}{r_{\rm ps}^{\,d-3}} > 0, \\[4pt]
 \partial_M \Phi_1
 &= -\frac{2\,r_{\rm isco}^{\,1-4d}}{(d-1)(d-2)}\,
 \sqrt{h(r_{\rm isco},q,P,d)} < 0,
 \end{aligned}
 \label{eq:p0p1}
\end{equation}
where the explicit expression of $h$ is omitted for brevity. Substituting eq.~\eqref{eq:p0p1} into the criterion~\eqref{eq:pj1} and using eqs.~\eqref{eq:g0g1} and \eqref{eq:dm1}, we find
\begin{align}
 \epsilon = 0:&\quad
 -\frac{1}{\Gamma_0}\frac{\dd M}{\dd r_h} > 0,\
 \partial_M \Phi_0 > 0
 \ \Rightarrow\
 \frac{\dd r_{\rm ps}}{\dd r_h} > 0, \\[2pt]
 \epsilon = 1:&\quad
 -\frac{1}{\Gamma_1}\frac{\dd M}{\dd r_h} < 0,\
 \partial_M \Phi_1 < 0
 \ \Rightarrow\
 \frac{\dd r_{\rm isco}}{\dd r_h} > 0.
\end{align}
Hence, along the isobaric process, both $r_{\rm ps}(r_h)$ and $r_{\rm isco}(r_h)$ increase monotonically. The opposite signs of $\Gamma_0$ and $\Gamma_1$ are exactly compensated by those of $\partial_M \Phi_0$ and $\partial_M \Phi_1$, leading to the same monotonic behavior of $r_{\rm ps}(r_h)$ and $r_{\rm isco}(r_h)$. These findings are consistent with the earlier results reported by Wei, Zhang, et al.~\cite{Wei:2017mwc,Zhang:2019tzi}.

Next, we take $d=4$ as an explicit example and compute the expression of ${\dd r_{\rm c}}/{\dd r_h}$. For the four-dimensional RN--AdS black hole, one obtains
\begin{align}
 \frac{\dd M}{\dd r_h}&=\frac{2\left(-{q^{2}}/{r_{h}^{3}} + {1}/{r_{h}} + 8\pi P\,r_{h}\right)}{{c_d}/{r_h^{\,d-3}}}>0, \notag\\[4pt]
 \Gamma_0 &= \frac{4 q^{2} - 2 r_{\rm ps}^{2}}{r_{\rm ps}^{3}}<0, \notag\\[4pt]
 \Gamma_1 &= \frac{1}{6\,r_{\rm isco}^{6}}
 \Bigl[
 15 q^{4} + r_{\rm isco}^{4}
 + 352\pi P\, r_{\rm isco}^{6}
 + 4800\pi^{2} P^{2} r_{\rm isco}^{8}
 +\, r_{\rm isco}^{2}\sqrt{g} \notag\\
 &\qquad \qquad \qquad \qquad \quad
 - 120\pi P\, r_{\rm isco}^{4}\sqrt{g}
 +\, q^{2}\!\left(48\pi P\, r_{\rm isco}^{4} - 9\sqrt{g}\right)
 \Bigr]>0,
 \label{eq:p0p12}
\end{align}
where
\begin{equation}
 g =-15 q^4 + r_{\rm isco}^4
 + 176\pi P\,r_{\rm isco}^6
 + 1600\pi^2 P^2 r_{\rm isco}^8
 + 6 q^2 r_{\rm isco}^2\,\left(3+8\pi P\,r_{\rm isco}^2\right).
\end{equation}
Using the criterion~\eqref{eq:pj1} together with the above results, we find
\begin{equation}
 \frac{\dd r_{\rm ps}}{\dd r_h}
 =\frac{3\,r_{\rm ps}^2\!\left(q^2 - r_h^2 - 8\pi P\,r_h^4\right)}
 {2\left(2 q^2 - r_{\rm ps}^2\right) r_h^2},
 \label{eq:dyrnads1}
\end{equation}
and
\begin{multline}
\frac{\dd r_{\rm isco}}{\dd r_h}
=-\,6\,\left(q^2 - r_h^2 - 8\pi P\,r_h^4\right)\,r_{\rm isco}^{13}\,\sqrt{g}\;/\\
\Bigl[ r_h^2\Bigl( 15 q^4 r_{\rm isco}^{11}
+ r_{\rm isco}^{15}
+ 352\pi P\, r_{\rm isco}^{17}
+ 4800\pi^2 P^2 r_{\rm isco}^{19} \\
+ 48\pi P\, q^2\, r_{\rm isco}^{15}
+\left(r_{\rm isco}^2 - 120\pi P\, r_{\rm isco}^4 - 9 q^2\right)\,r_{\rm isco}^{11}\,\sqrt{g}
\Bigr)
\Bigr].
\label{eq:dyrnads2}
\end{multline}
In the Sch--AdS limit ($q=0$), the above relations reduce to
\begin{align}
 \frac{\dd r_{\rm ps}}{\dd r_h}&=\frac{3}{2}+12 \pi P r_h^2,
 \label{eq:dysads1}\\
 \frac{\dd r_{\rm isco}}{\dd r_h}
 &=\frac{6\,\left(1 + 8\pi P\,r_h^2\right)\,
 \sqrt{j}}{1 + 4800\pi^2 P^2 r_{\mathrm{isco}}^4 + \sqrt{j}
 - 8\pi P\,r_{\mathrm{isco}}^2
 \left(-44 + 15\sqrt{j}\right)},
 \label{eq:dysads2}
\end{align}
where $j = 1 + 176\pi P\,r_{\mathrm{isco}}^2 + 1600\pi^2 P^2 r_{\mathrm{isco}}^4$. Further, in the limit where both $q$ and $P$ approach zero, one obtains
\begin{equation}
 \frac{\dd r_{\rm ps}}{\dd r_h}=\frac{3}{2},
 \qquad
 \frac{\dd r_{\rm isco}}{\dd r_h}=3,
 \label{eq:dys}
\end{equation}
which are consistent with the well-known relations $r_{\rm ps}=3M$, $r_{\rm isco}=6M$, and $r_h=2M$ in the Schwarzschild case.

\begin{figure}[htbp]
 \centering
 \begin{minipage}{0.48\linewidth}
 \centering
 \includegraphics[width=\linewidth]{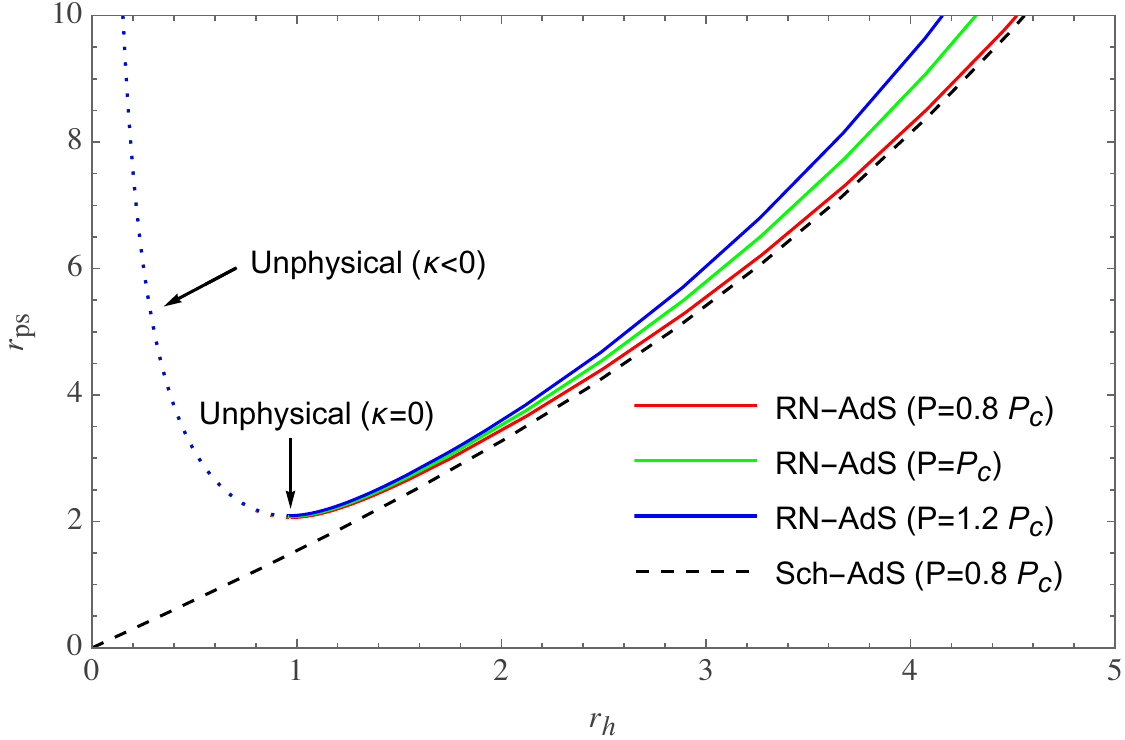}
 \end{minipage}
 \hfill
 \begin{minipage}{0.48\linewidth}
 \centering
 \includegraphics[width=\linewidth]{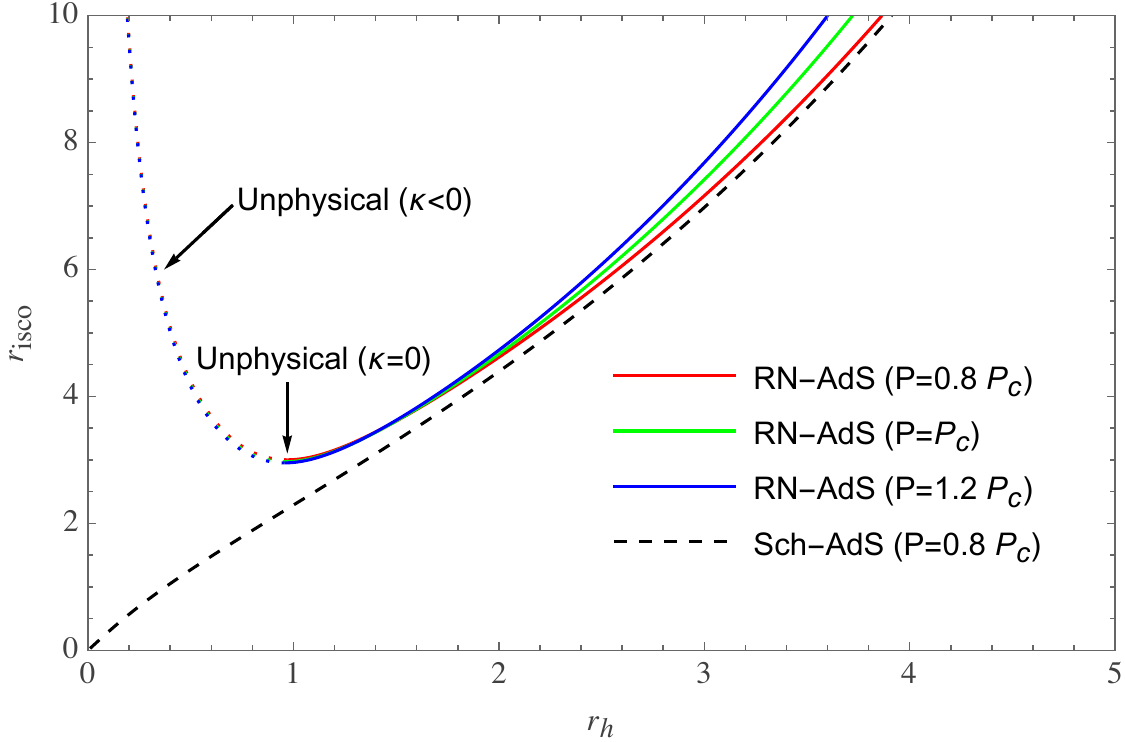}
 \end{minipage}

 \caption{Curves of $r_{\rm ps}(r_h)$ (left panel) and $r_{\rm isco}(r_h)$ (right panel) along isobars for $d=4$ and $q=1$. For the Sch--AdS case, the pressure is likewise set to the value $P = 0.8\,P_{\mathrm{c}}$, where $P_{\mathrm{c}}$ is the critical pressure of the RN--AdS solution with $q=1$.
 }
 \label{dyrpsrisco}
\end{figure}

In figure~\ref{dyrpsrisco}, we present the numerically computed profiles of $r_{\mathrm{ps}}(r_h)$ and $r_{\mathrm{isco}}(r_h)$ for the four-dimensional case. As illustrated, $r_{\mathrm{ps}}(r_h)$ and $r_{\mathrm{isco}}(r_h)$ remain globally monotonically increasing for the Sch--AdS black hole, which is consistent with our expectations. However, for the RN--AdS black hole, they first decrease and then increase with $r_h$, exhibiting nonmonotonic behavior. In fact, they should be truncated at the minimum, and only the increasing branch should be kept. This is because the minimum occurs exactly at the minimum of $M(r_h)$. According to eq.~\eqref{eq:dm1}, the temperature there is $T=0$, or equivalently the surface gravity is $\kappa=2\pi T=0$, which is unphysical and should be discarded. After truncation, $r_{\mathrm{ps}}(r_h)$ and $r_{\mathrm{isco}}(r_h)$ are both monotonically increasing, which is consistent with our expectations.

\subsection{Isothermal process}

Along the isothermal process, only the charge $q$ in the metric~\eqref{eq:rnads} is held fixed, and the thermodynamic evolution is jointly driven by variations of the mass $M$ and the pressure $P$. This corresponds to the case in section~\ref{sec22} where both $M$ and $P$ vary. In this case, the criterion~\eqref{eq:zy1} takes its full form,
\begin{equation}
 \frac{\dd r_{\rm c}}{\dd r_h}
 = -\,\frac{1}{\Gamma_\epsilon}\!\left(
 \frac{\dd M}{\dd r_h}\,\partial_M\Phi_\epsilon
 +\frac{\dd P}{\dd r_h}\,\partial_P\Phi_\epsilon
 \right).
 \label{eq:pj2}
\end{equation}
For the metric~\eqref{eq:rnads}, one readily verifies that $\partial_P \Phi_{0}\equiv 0$. This corresponds to case~(i) in section~\ref{sec22}, so $r_{\rm ps}(r_h)$ can be nonmonotonic. Moreover, eq.~\eqref{eq:p0p1} shows that $\partial_M \Phi_{1}<0$, and Appendix~\ref{appd} proves that $\partial_P \Phi_{1}>0$. These two derivatives therefore have opposite signs, which matches case~(ii) in section~\ref{sec22} and implies that $r_{\rm isco}(r_h)$ can be nonmonotonic. We next confirm that such nonmonotonic behavior indeed occurs.

\subsubsection{Photon sphere ($\epsilon=0$)}

For the photon sphere, $\partial_{P}\Phi_{0}\equiv 0$, and the criterion reduces to
\begin{equation}
 \frac{\dd r_{\mathrm{c}}}{\dd r_h}
 = -\,\frac{1}{\Gamma_0}\,
 \frac{\dd M}{\dd r_h}\,
 \partial_{M}\Phi_0.
 \label{eq:pj3}
\end{equation}
This has the same form as in the isobaric case, but here the first law allows ${\dd M}/{\dd r_h}$ to cross zero. For the metric~\eqref{eq:rnads} one obtains
\begin{equation}
 \begin{aligned}
 \frac{\dd M}{\dd r_h}
 =\frac{1}{(d-1)\,c_d}\Big[
 2(d-3)\,r_h^{\,d-4}
 +4(d-2)\,\pi T_0\,r_h^{\,d-3}
 -2(d-2)(d-3)\,q^{2}\,r_h^{\,2-d}
 \Big],
 \end{aligned}
 \label{eq:dm2}
\end{equation}
and
\begin{equation}
 \Gamma_0=\frac{2(d-3)}{r_{\rm ps}}\Big[(d-2)\,q^{2}\,r_{\rm ps}^{\,6-2d}-1\Big],
\end{equation}
where $T_0$ denotes the temperature along the isothermal process. Moreover, $\partial_M \Phi_0$ is given by eq.~\eqref{eq:p0p1}. We now analyze the sign of ${\dd M}/{\dd r_h}$. For $q=0$ one immediately has ${\dd M}/{\dd r_h}>0$. For $q\neq0$, differentiating ${\dd M}/{\dd r_h}$ gives
\begin{equation}
 \begin{aligned}
 \frac{\dd ^2M}{\dd r_h^2}
 =\frac{1}{(d-1)c_d}\Big[\,2(d-3)(d-4)\,r_h^{\,d-5}
 +4(d-2)(d-3)\pi T_0\,r_h^{\,d-4}
 +2(d-2)^2(d-3)q^2\,r_h^{\,1-d}\Big]>0,
 \end{aligned}
\end{equation}
hence ${\dd M}/{\dd r_h}$ is strictly increasing in $r_h$. Meanwhile, eq.~\eqref{eq:dm2} implies
\begin{equation}
 \lim_{r_h\to0^+}\frac{\dd M}{\dd r_h}=-\infty,
 \qquad
 \lim_{r_h\to+\infty}\frac{\dd M}{\dd r_h}=+\infty,
\end{equation}
hence there exists a unique $r_{h,e_1}$ such that $\bigl.{\dd M}/{\dd r_h}\bigr|_{r_{h,e_1}}=0$. Therefore, by applying the criterion~\eqref{eq:pj3}, we deduce the following behavior for the photon sphere radius $r_{\rm ps}(r_h)$ as a function of the horizon radius $r_h$:
\begin{itemize}
\item Case $q=0$: $r_{\rm ps}(r_h)$ is monotonically increasing;
\item Case $q\neq0$: $r_{\rm ps}(r_h)$ is nonmonotonic. It decreases for small $r_h$, reaches a single minimum at $r_{h,e_1}$, and then increases.
\end{itemize}
These results confirm our previous analysis that $r_{\rm ps}(r_h)$ can indeed become nonmonotonic, which has not been noticed in earlier studies. Interestingly, in the Sch--AdS case ($q=0$), $r_{\rm ps}(r_h)$ remains monotonic. This can be readily understood from eq.~\eqref{eq:zy2}: when $q=0$, the slope $\mathrm{d}P/\mathrm{d}r_h$ is not sufficiently negative to drive $\mathrm{d}M/\mathrm{d}r_h$ across zero. By contrast, for the RN--AdS case ($q\neq 0$), $P(r_h)$ exhibits a Van der Waals-like behavior, and at small $r_h$ its slope $\mathrm{d}P/\mathrm{d}r_h$ becomes strongly negative, which is sufficient to make $\mathrm{d}M/\mathrm{d}r_h$ cross zero.

The explicit expression for ${\dd r_{\mathrm{ps}}}/{\dd r_h}$ reads
\begin{equation}
 \frac{\dd r_{\rm ps}}{\dd r_h}=\frac{\Big[
 (d-3)(2-d)\,q^2\,r_h^{6}+\bigl(d-3+2(d-2)\pi T_0\,r_h\bigr)r_h^{2d}\Big]\,
 r_{\rm ps}^{\,4+d}}{(d-3)\Bigl[(2-d)\,q^2\,r_{\rm ps}^{6}+r_{\rm ps}^{2d}\Bigr]\,r_h^{\,4+d}}.
\end{equation}
For $d=4$ (the RN--AdS case), this reduces to
\begin{equation}
 \frac{\dd r_{\rm ps}}{\dd r_h}
 =\frac{r_{\rm ps}^{2}\,\left[2 q^{2} - r_h^{2}\,\left(1 + 4\pi T_0\, r_h\right)\right]}
 {(2 q^{2} - r_{\rm ps}^{2})\, r_h^{2}},
\end{equation}
and, in the Sch--AdS limit ($q=0$), further simplifies to
\begin{equation}
 \frac{\dd r_{\rm ps}}{\dd r_h}= 1 + 4\pi T_0\,r_h.
 \label{eq:dtsads}
\end{equation}
This coincides with the standard Sch--AdS result, from which ${\dd r_{\rm ps}}/{\dd r_h}>0$ is immediately evident.

\begin{figure}[htbp]
 \centering
 \includegraphics[width=0.48\linewidth]{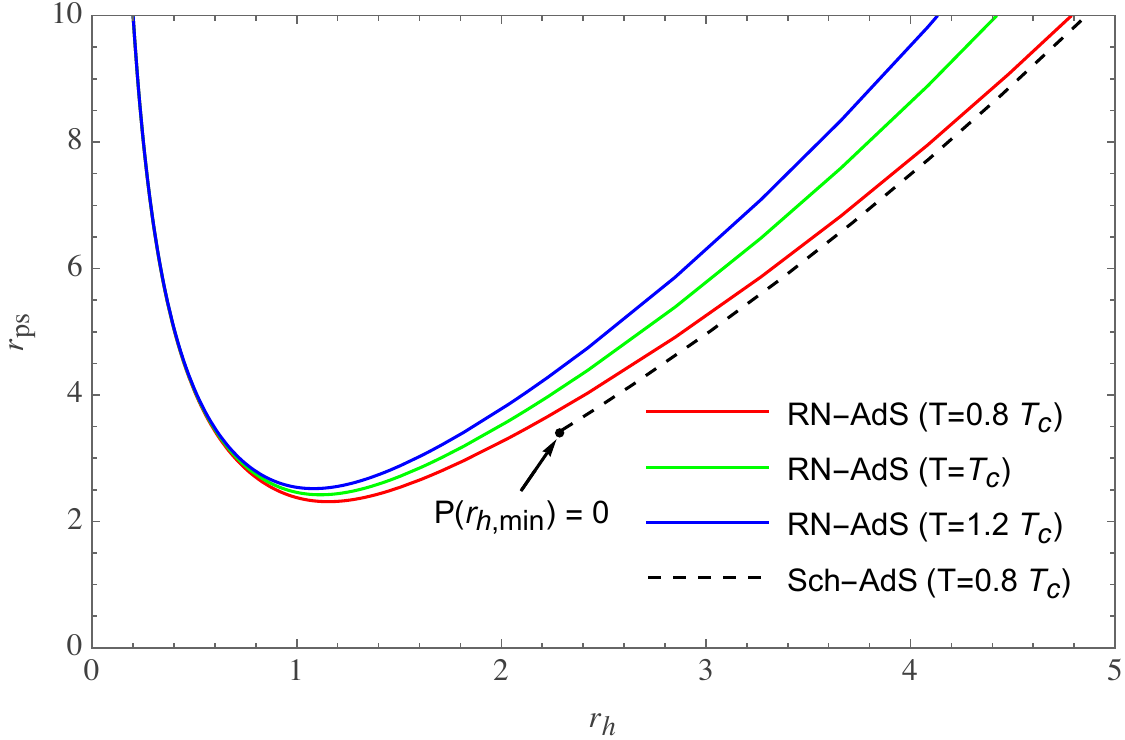}

 \caption{Curves of $r_{\rm ps}(r_h)$ along isotherms for $d=4$ and $q=1$. For the Sch--AdS case, we likewise set the temperature to $T = 0.8\,T_{\mathrm{c}}$, where $T_{\mathrm{c}}$ is the critical temperature of the RN--AdS solution with $q=1$. The same choice is adopted for subsequent figures and will not be repeated.}
 \label{dtrps}
\end{figure}

To verify the above analysis, we have performed numerical calculations and displayed the results in figure~\ref{dtrps}. For the Sch--AdS black hole, $r_{\mathrm{ps}}(r_h)$ begins at $r_{h,\min}=(d-3)/(4\pi T_0)$, where $r_{h,\min}$ is defined by $P(r_{h,\min})=0$, and smaller $r_h$ would give $P<0$ and are therefore discarded as unphysical. It then increases monotonically, remaining globally monotonic in full agreement with eq.~\eqref{eq:dtsads}. In contrast, the RN--AdS black hole exhibits the expected nonmonotonic behavior: although ${\mathrm{d}M}/{\mathrm{d}r_h}=0$ at the extremum, the surface gravity remains positive ($\kappa = 2\pi T_0 > 0$), indicating that the black hole is nonextremal and the curve need not be truncated.

\subsubsection{ISCO ($\epsilon=1$)}

The metric~\eqref{eq:rnads} yields
\begin{equation}
 \begin{aligned}
 \frac{\mathrm{d}P}{\mathrm{d}r_h}
 = -\frac{(d-2)}{8\pi}
 \Bigl[
 (d-2)(d-3)\,q^2\,r_h^{\,3-2d}
 + \bigl(3-d+2\pi T_0 r_h\bigr)\,r_h^{-3}
 \Bigr],
 \end{aligned}
 \label{eq:dp}
\end{equation}
and
\begin{equation}
 \begin{aligned}
 \partial_P\Phi_1
 = \frac{16\pi\, r_{\mathrm{isco}}}{(d-2)(d-1)}
 \Bigl[\,8 + 4(d-1)(d-2)q^{2}r_{\mathrm{isco}}^{6-2d}
 - c_{d}(d-1)(d+1)M\,r_{\mathrm{isco}}^{3-d}\Bigr],
 \end{aligned}
\end{equation}
where $\partial_P\Phi_1>0$ (see Appendix~\ref{appd} for proof). Since van der Waals-like behavior occurs only for charged black holes, we analyze the RN--AdS and Sch--AdS cases separately. Moreover, because the explicit expression for ${\dd r_{\mathrm{c}}}/{\dd r_h}$ is algebraically cumbersome, we focus on whether it changes sign rather than displaying the full formula.

\subsubsection*{(i) Sch--AdS ($q=0$)}

For $q=0$, eq.~\eqref{eq:dp} reduces to
\begin{equation}
 \frac{\mathrm{d}P}{\mathrm{d}r_h}
 = -\,\frac{(d-2)\,\bigl(3-d+2\pi T_0 r_h\bigr)}{8\pi\, r_h^{3}}.
\end{equation}
It has a single zero at $r_{h,e_2}={(d-3)}/{2\pi T_0}$, with ${\mathrm{d}P}/{\mathrm{d}r_h}>0$ for $r_h<r_{h,e_2}$ and ${\mathrm{d}P}/{\mathrm{d}r_h}<0$ for $r_h>r_{h,e_2}$. Using the criterion~\eqref{eq:pj2}, together with ${\dd M}/{\dd r_h}\,\partial_M\Phi_1<0$, $\Gamma_1>0$, and $\partial_P\Phi_1>0$, one thus expects $r_{\rm isco}(r_h)$ to first decrease and then increase. To make this more explicit, we rewrite the criterion~\eqref{eq:pj2} as
\begin{equation}
 \frac{\dd r_{\rm c}}{\dd r_h}
 =-\,\frac{1}{\Gamma_\epsilon}\,\delta(r_{\rm isco},r_h,d,q,T_0).
\end{equation}
For sufficiently large $r_h$ one has ${\mathrm{d}P}/{\mathrm{d}r_h}<0$, which forces $\delta<0$. It remains to verify that $\delta>0$ for small $r_h$, ensuring that $\delta$ crosses zero at least once. In this spirit we find
\begin{equation}
 \begin{cases}
 \delta\bigl(3r_{h,\min},\,r_{h,\min},\,4,\,0,\,T_0\bigr)
 =\dfrac{1232\,\pi^2 T_0^2}{27}>0, & d=4,\\[6pt]
 \delta\bigl(+\infty,\,r_{h,\min},\,d,\,0,\,T_0\bigr)
 =\dfrac{512\,\pi^3 T_0^3\,r_{\rm isco}}
 {(d-3)^{2}(d-1)}>0, & d>4,
 \end{cases}
 \label{eq:dd}
\end{equation}
where $r_{h,\min}=(d-3)/(4\pi T_0)$ is defined by $P(r_{h,\min})=0$. For $d=4$ and $P=0$, the metric~\eqref{eq:rnads} reduces to the Schwarzschild case, giving $r_{\rm isco}=3\,r_{h,\min}$. For $d>4$ and $P=0$, the metric~\eqref{eq:rnads} reduces to the higher-dimensional Schwarzschild solution, which admits no ISCO. This corresponds to $\lim_{P\to0} r_{\rm isco}=+\infty$, see the end of Appendix~\ref{appd} for a brief explanation. Equation~\eqref{eq:dd} shows that $\delta>0$ at small $r_h$, and together with $\delta<0$ at sufficiently large $r_h$ this guarantees that $\delta$ crosses zero, implying that $r_{\rm isco}(r_h)$ is necessarily nonmonotonic on its full domain.

\subsubsection*{(ii) RN--AdS ($q\neq 0$)}

Since the RN--AdS black hole exhibits a van der Waals-like phase transition, one has
\begin{equation}
 \frac{\dd P}{\dd r_h}\le 0 \quad \text{for}\,\,T_0\ge T_c,
\end{equation}
where the equality holds precisely at the critical point $(r_{h,c},P_c,T_c)$. According to the criterion~\eqref{eq:pj2}, when $r_h\ge r_{h,e_1}$ (where $r_{h,e_1}$ is the zero of $\dd M/\dd r_h$), it follows that $\delta<0$ and hence $\dd r_{\mathrm{isco}}/\dd r_h>0$, so $r_{\mathrm{isco}}(r_h)$ increases monotonically. For $r_h<r_{h,e_1}$, however, $\delta$ may become positive. To determine this behavior, we examine the limit $r_h\to0^+$.

From eq.~\eqref{eq:zy2}, one obtains
\begin{equation}
 \lim_{r_h\to0^+}\frac{\dd M}{\dd r_h}
 =V\,\frac{\dd P}{\dd r_h},
 \label{eq:dm_smallrh}
\end{equation}
which substituted into the criterion~\eqref{eq:pj2} gives
\begin{equation}
 \delta
 =\frac{\dd P}{\dd r_h}\!\left(V\,\partial_M\Phi_1+\partial_P\Phi_1\right).
 \label{eq:delta_expr}
\end{equation}
Meanwhile, from Appendix~\ref{appe}, as $r_h\to0^+$ one has
\begin{equation}
 V\,\partial_M\Phi_1\ \longrightarrow\ -\infty,
 \qquad
 \partial_P\Phi_1\ \longrightarrow\ 0.
 \label{eq:phi_limits}
\end{equation}
Hence,
\begin{equation}
 \delta
 =\frac{\dd P}{\dd r_h}\Bigl(V\,\partial_M\Phi_1+\partial_P\Phi_1\Bigr)>0,
 \quad\Rightarrow\quad
 \frac{\dd r_{\mathrm{isco}}}{\dd r_h}<0.
 \label{eq:final_delta}
\end{equation}
This demonstrates that $r_{\mathrm{isco}}(r_h)$ is nonmonotonic over the full domain of $r_h$.

The same conclusion holds for $T_0<T_c$. Along both the small black hole and large black hole stable branches, one still has $\dd P/\dd r_h<0$, and the leading-order behavior as $r_h\to0^+$ remains identical to the $T_0\ge T_c$ case above. The only difference is that for $T_0<T_c$ an intermediate spinodal (unstable) branch appears, on which $\dd P/\dd r_h>0$. Along this segment $\delta$ could, in principle, change sign. However, we consider this possibility to be extremely unlikely. In addition, this unstable branch is dynamically skipped by the first-order phase transition, so it can be safely ignored. In any case, viewed globally, $r_{\rm isco}(r_h)$ is necessarily nonmonotonic for every fixed $T_0$. These results again confirm our previous analysis that $r_{\rm isco}(r_h)$ can indeed exhibit nonmonotonic behavior.

\begin{figure}[htbp]
 \centering
 \includegraphics[width=0.48\linewidth]{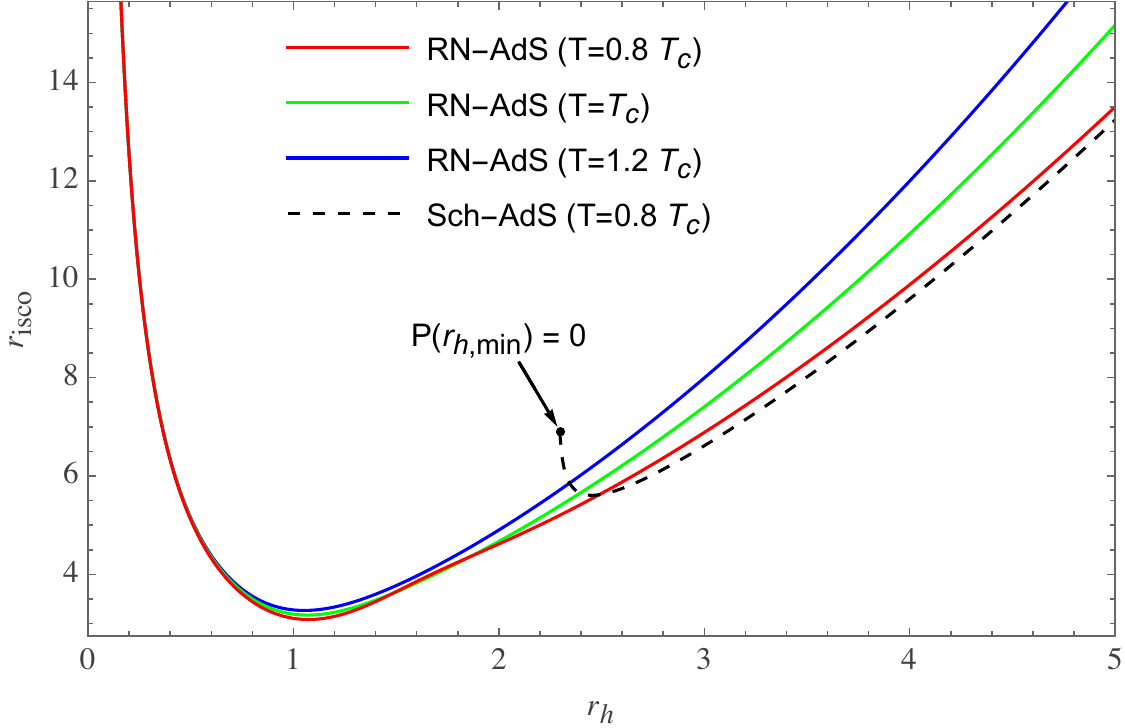}

 \caption{Curves of $r_{\rm isco}(r_h)$ along isotherms for $d=4$ and $q=1$.}
 \label{dtrisco}
\end{figure}

To verify the above analysis, we perform the numerical check. The result is displayed in figure~\ref{dtrisco}. It can be seen that $r_{\mathrm{isco}}(r_h)$ is nonmonotonic for both charged and uncharged cases, in full agreement with our above analysis.

At this stage, we have completed the analysis of the behavior of $r_c(r_h)$ along isobaric and isothermal processes. As anticipated, $r_c(r_h)$ can be either monotonic or nonmonotonic, and this behavior is closely tied to the first law. Next, taking the four-dimensional RN--AdS black hole as an example, we further verify that even though $r_c(r_h)$ is nonmonotonic along isotherms, the jump $\Delta r_c = r_c(r_{h,l})-r_c(r_{h,s})$ remains nonzero. Moreover, for the $d$-dimensional AdS black holes considered here, eqs.~\eqref{eq:p0p1} and \eqref{eq:p0p12} imply
\begin{equation}
 \frac{1}{\Gamma_\epsilon}\,\partial_M\Phi_\epsilon<0.
\end{equation}
Meanwhile, along the isothermal process, the entropy increases across the transition, and eq.~\eqref{eq:ddm} then gives $M_l>M_s$. Substituting these sign relations into eq.~\eqref{eq:drc}, one readily finds $\Delta r_c>0$. We therefore expect $r_c(r_{h,l})>r_c(r_{h,s})$. Finally, it should be pointed out that the Sch--AdS black hole does not exhibit a small/large black hole phase transition, but only the Hawking-Page phase transition. This can be viewed as a transition from a thermal AdS phase with no black hole to a black hole phase, so $r_c$ must jump together with $r_h$ (from no $r_c$ to a finite $r_c$). Therefore, we do not need to verify the Sch--AdS case.

\subsection{Verifying the jump in RN--AdS}

Our verification strategy is simple: we numerically scan all varying parameters to obtain $r_{h,s}$ and $r_{h,l}$, and then substitute them into $r_c(r_h)$ to compare $r_c(r_{h,s})$ and $r_c(r_{h,l})$. For the RN--AdS black hole in the isothermal process, the varying parameters are the temperature $T_0$ and the charge $q$. Conveniently, the explicit $q$-dependence can be removed by introducing reduced (dimensionless) variables
\begin{equation}
 \tilde r_{h}=\frac{r_{h}}{r_{h,c}},\qquad
 \tilde P=\frac{P}{P_{c}},\qquad
 \tilde T=\frac{T}{T_{c}},
\end{equation}
where $r_{h,c}$, $P_c$, and $T_c$ are given by eq.~\eqref{eq:rhpt}. Using these reduced variables, the isobaric relation in eq.~\eqref{eq:prh} takes the form
\begin{align}
 \tilde P(\tilde r_h,\tilde T_0)
 = \frac{1 - 6\tilde r_h^{2} + 8\tilde r_h^{3}\tilde T_0}{3\tilde r_h^{4}},
 \label{eq:wlgp}
\end{align}
which no longer contain $q$ explicitly, so the conclusions derived from them apply to arbitrary charge. Similarly, $r_{\rm ps}(r_h)$ and $r_{\rm isco}(r_h)$ can be written in the reduced parameter space. It should be noted that only $r_{\rm ps}(r_h)$ has a closed-form expression, which reduces to
\begin{equation}
 \begin{aligned}
 \tilde r_{\rm ps}(\tilde r_{h})
 =\frac{1}{2\left(3+\sqrt{6}\right)\tilde r_h} \Bigg[1+\left(3+2\tilde T_0\tilde r_h\right)\tilde r_h^2
 +\sqrt{1-6\tilde r_h^{2}+4\tilde T_0 \tilde r_h^{3}+\left(3+2\tilde T_0\tilde r_h\right)^{2}\tilde r_h^{4}}\,\Bigg].
 \end{aligned}
 \label{eq:wlgrps}
\end{equation}
Although $r_{\rm isco}(r_h)$ has no closed-form expression, rewriting the ISCO conditions~\eqref{eq:phi1} in the reduced parameter space allows one to compute $\tilde r_{\rm isco}(\tilde r_h)$ numerically. It is likewise independent of $q$. Now, the only remaining control parameter is $\tilde T_0$. It is worth noting that to ensure $\tilde P(\tilde r_h,\tilde T_0)\ge0$ for all $\tilde r_h$, one must have $\tilde T_0\ge\sqrt{2}/2$, while the first-order phase transition occurs only for $T_0<T_c$, i.e., $\tilde T_0<1$. Thus, the physically relevant range is $\tilde T_0\in[\sqrt{2}/2,\,1)$.

\begin{figure}[htbp]
 \centering
 \begin{minipage}{0.48\linewidth}
 \centering
 \includegraphics[width=\linewidth]{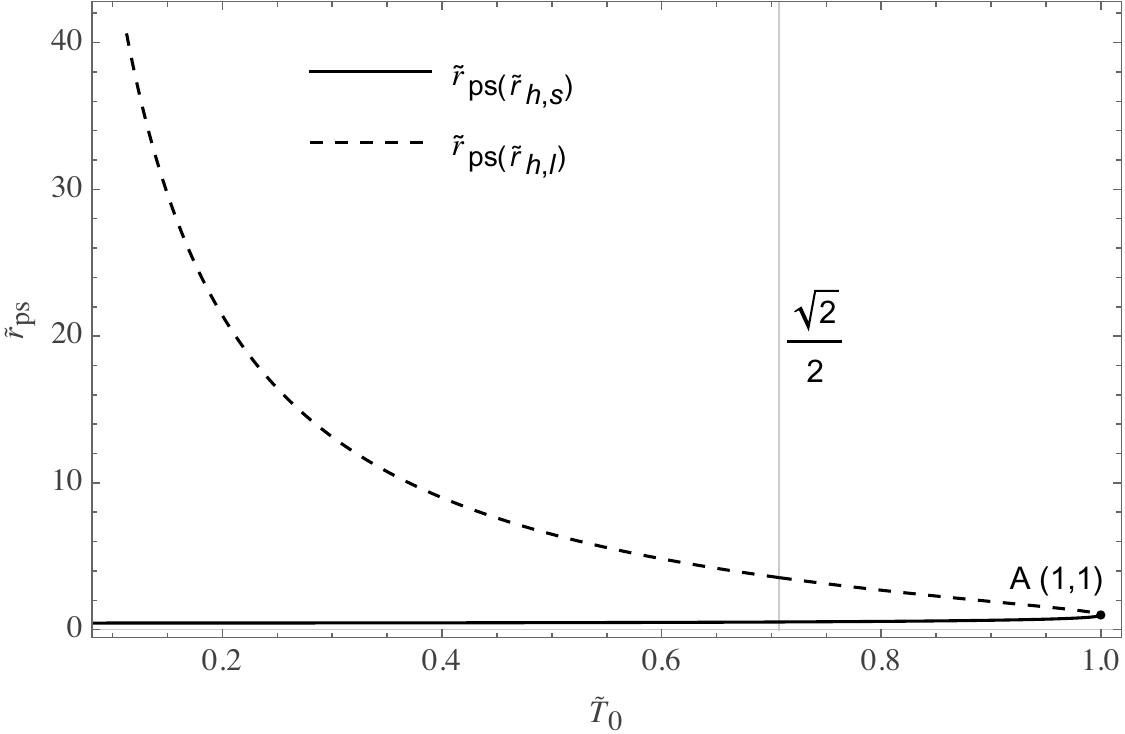}
 \end{minipage}
 \hfill
 \begin{minipage}{0.48\linewidth}
 \centering
 \includegraphics[width=\linewidth]{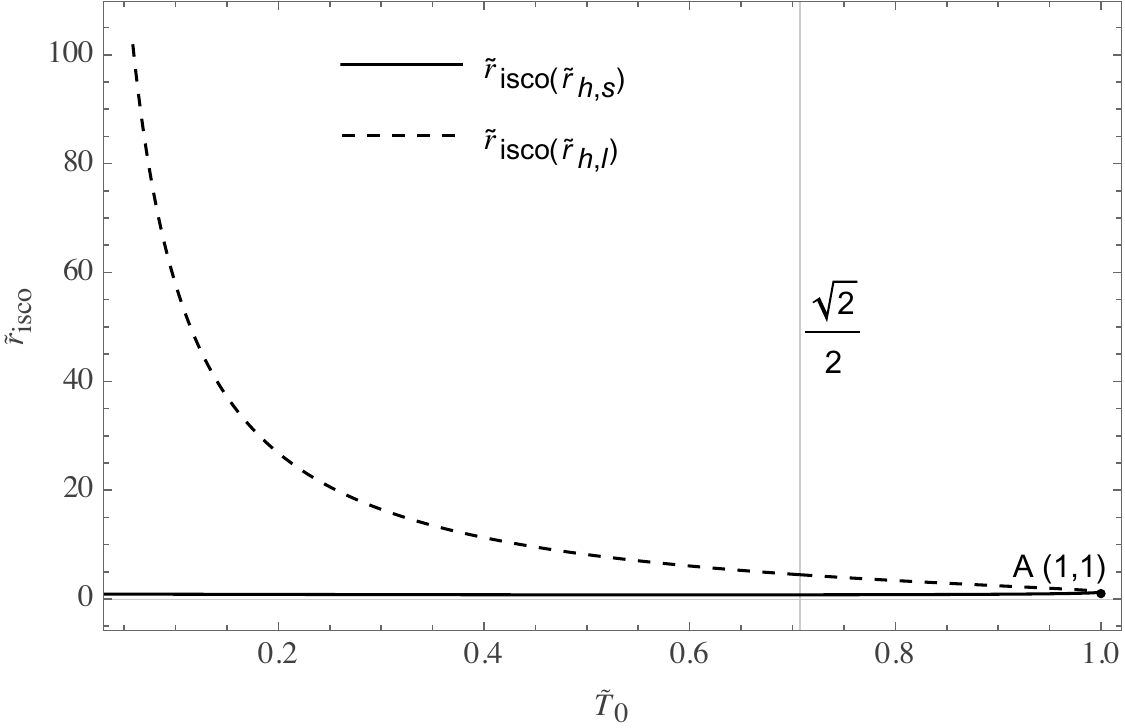}
 \end{minipage}

 \caption{Curves of $\tilde r_{\rm ps}(\tilde r_{h,s})$, $\tilde r_{\rm ps}(\tilde r_{h,l})$ (left panel), and $\tilde r_{\rm isco}(\tilde r_{h,s})$, $\tilde r_{\rm isco}(\tilde r_{h,l})$ (right panel) as functions of $\tilde T_0$. In both panels, the solid and dashed curves intersect at point $A$, where $r_{h,s}=r_{h,l}=r_{h,c}$. Note that only the region with $\tilde T_0 \ge \sqrt{2}/2$ is physical, whereas the region with $\tilde T_0<\sqrt{2}/2$ is unphysical.}
 \label{rpsiscot}
\end{figure}

Next, we only need to compute $\tilde r_{h,s}$ and $\tilde r_{h,l}$ numerically from eq.~\eqref{eq:wlgp} via the Maxwell equal-area construction, and then substitute them into eq.~\eqref{eq:wlgrps} and the numerical function $\tilde r_{\rm isco}(\tilde r_h)$. The result is displayed in figure~\ref{rpsiscot}. The solid and dashed black curves correspond to $\tilde r_c(\tilde r_{h,s})$ and $\tilde r_c(\tilde r_{h,l})$, respectively. From these figures, it is clear that $\tilde r_c(\tilde r_{h,l})\ge \tilde r_c(\tilde r_{h,s})$. The equality holds only at point A, which corresponds to the critical point where $r_{h,s}=r_{h,l}=r_{h,c}$. Therefore, except at the critical point, whenever $r_h$ jumps, $r_c$ always jumps with it and always to a larger value. This fully agrees with our expectation and with the conclusion of section~\ref{sec22}.

\section{Critical exponent invariance}
\label{sec4}

In the previous two sections, we proved and verified that $r_c$ jumps together with $r_h$ at a phase transition. This naturally raises a question: if we take $\Delta r_c= r_{c,l}-r_{c,s}$ as an order parameter, does it exhibit the same universal critical behavior as $\Delta r_h\equiv r_{h,l}-r_{h,s}$? Specifically, near the critical point (e.g., at the open circles in figure~\ref{pt}), the distinction between the coexisting phases becomes progressively weaker and vanishes exactly at the critical point. In conventional thermodynamics, such critical behavior is found to be universal and can be quantitatively characterized by critical exponents. For example, the classical Van der Waals system exhibits the order-parameter critical exponent $\beta=1/2$, which has also been shown to hold for many AdS black holes when the thermodynamic volume or the horizon radius is adopted to define the order parameter. Recently, Wei et al. demonstrated that in the isobaric process the four-dimensional RN--AdS black hole yields the same value $\beta=1/2$ when the photon sphere radius is used to define the order parameter, and their numerical results for higher dimensional RN--AdS black holes remain very close to $1/2$~\cite{Wei:2017mwc}. These findings motivate us to test whether $\Delta r_c$ retains a universal critical exponent in a broader class of black hole spacetimes. Next, we will prove that the mapping $r_c (r_h)$ is necessarily locally monotonic in a neighborhood of the critical point. As a result, the reparametrization $r_h\mapsto r_c $ reduces to a smooth linear transformation in a neighborhood of the critical point. This ensures that $\Delta r_c$ yields the same critical exponent as $\Delta r_h$.

The order-parameter critical exponent $\beta$ is defined from the coexistence gap of the horizon radius for $T<T_c$:
\begin{equation}
 \Delta r_h = r_{h,l}-r_{h,s} \sim |t|^{\beta},
 \qquad
 t:=\frac{T-T_c}{T_c}\to 0^-.
 \label{eq:rht}
\end{equation}
We will show that the circular orbit radius inherits the same scaling, namely
\begin{equation}\label{eq:deltarc}
 \Delta r_c = r_c\big(r_{h,l}\big)-r_c\big(r_{h,s}\big)\sim |t|^{\beta}.
\end{equation}

We first expand $r_c(r_h)$ in the vicinity of the critical point as
\begin{equation}
 r_c(r_h)
 = r_c(r_{h,c})
 + \left.\frac{\dd r_c}{\dd r_h}\right|_{\rm c}\,\bigl(r_h-r_{h,c}\bigr)
 + o\left(|r_h-r_{h,c}|\right),
\end{equation}
where $r_{h,c}$ is the horizon radius evaluated at the critical point and the subscript ``$\left.\right|_{\rm c}$'' denotes evaluation at the critical point. From this expansion, one obtains
\begin{equation}
 \Delta r_c
 = r_c\big(r_{h,l}\big)-r_c\big(r_{h,s}\big)
 = \left.\frac{\dd r_c}{\dd r_h}\right|_{\rm c}\,\Delta r_h + o(\Delta r_h).
\end{equation}
This relation implies that as long as $\left.\dd r_c/\dd r_h\right|_{\rm c}\neq 0$, the leading behavior of $\Delta r_c$ in the limit $t\to 0^{-}$ is controlled by $\Delta r_h$. In other words, they generate the same critical exponent.

Next, we examine whether $\left.\dd r_c/\dd r_h\right|_{\rm c}$ can vanish. For a black hole system exhibiting a second-order phase transition, the equation of state satisfies
\begin{equation}
 \left.\frac{\partial P}{\partial r_h}\right|_{\rm c}
 =\left.\frac{\partial^2 P}{\partial r_h^2}\right|_{\rm c}=0
\end{equation}
at the critical point. Substituting these conditions into eqs.~\eqref{eq:zy1} and \eqref{eq:zy2} yields
\begin{equation}
 \left.\frac{\dd r_c}{\dd r_h}\right|_{\rm c}
 = \left.-\,\frac{1}{\Gamma_\epsilon}\,
 \frac{\dd M}{\dd r_h}\,
 \partial_{M}\Phi_\epsilon \right|_{\rm c},
\end{equation}
and
\begin{equation}
 \left.\frac{\dd M}{\dd r_h}\right|_{\rm c}=\left. T\,S'(r_h)\right|_{\rm c}.
\end{equation}
Combining the above two equations yields
\begin{equation}
 \left.\frac{\dd r_c}{\dd r_h}\right|_{\rm c}
 = \left.-\,\frac{1}{\Gamma_\epsilon}\,\partial_{M}\Phi_\epsilon\,T\,S'(r_h) \right|_{\rm c},
\end{equation}
which is manifestly nonzero. Therefore, $\Delta r_c$ generates the same critical exponent as $\Delta r_h$, namely,
\begin{equation}
 \beta_{(r_c)}=\beta_{(r_h)}.
\end{equation}
This result is noteworthy: Although $r_c(r_h)$ may exhibit global nonmonotonic behavior, it must be monotonic in the neighborhood of the critical point, which ensures that the critical exponent remains invariant under reparametrization. This implies that, during black hole thermodynamic evolution, descriptions in terms of $r_c$ or $r_h$ fall into the same universality class. Moreover, it is worth emphasizing that this proof does not rely on the specific form of the metric~\eqref{eq:rnads}, but only on the criterion \eqref{eq:zy1} and the first law. The result therefore applies, in principle, to any static, spherically symmetric black hole with $\partial_{M}\Phi_\epsilon\neq 0$, which points to a more fundamental connection between black hole thermodynamics and spacetime geometry.

\section{Summary and discussion}
\label{sec5}

In this paper, we derived the monotonicity criterion~\eqref{eq:pj} that applies to arbitrary thermodynamic paths in the parameter space of any static, spherically symmetric black hole, thereby clarifying when the mapping between the horizon radius $r_h$ and the circular orbit radius $r_{\rm c}$ is monotonic. In this framework, the monotonicity of $r_{\rm c}(r_h)$ is governed by two factors: (i) the intrinsic geometric/orbital properties of the black hole and (ii) the chosen path in the extended thermodynamic parameter space. For any nonextremal black hole admitting an unstable photon sphere or an ISCO, the first factor does not by itself induce nonmonotonicity, whereas the second is decisive.

Under suitable simplifications and assumptions, we used criterion~\eqref{eq:pj} to examine how thermodynamic constraints along different paths affect the monotonicity of $r_c(r_h)$. We found that $r_c(r_h)$ must be monotonic when either the mass or the pressure is held fixed. By contrast, when neither the mass nor the pressure is fixed, $r_c(r_h)$ can become nonmonotonic. These constraints can be realized in physical processes, including isenthalpic, isobaric, and isothermal ones. For example, in our case of $d$-dimensional charged AdS black holes, isobars correspond to fixed pressure, while isotherms correspond to the case where both mass and pressure vary. As expected, $r_c(r_h)$ is monotonic along isobars but becomes nonmonotonic along isotherms. Notably, these results are closely tied to the first law of black hole thermodynamics, indicating an underlying link between black hole thermodynamics and spacetime geometry.

After showing that $r_c(r_h)$ can be nonmonotonic, a natural question is whether, in this case, $r_c$ may no longer follow the jump of $r_h$ at a phase transition, as illustrated in figure~\ref{mnm}(b). Previous studies focused on monotonic $r_c(r_h)$ and thus neither identified nor resolved this issue. The central goal of this work is to address it. We mainly focused on zeroth and first order phase transitions in the extended phase space. This covers many phase transitions with jumps of $r_h$, including van der Waals-like phase transitions, reentrant phase transitions, and triple-point behavior, as well as some novel zeroth-order small/large black hole phase transitions. Using the definitions of zeroth and first order phase transitions together with the differential form of criterion~\eqref{eq:pj}, we found the following results. For a first-order phase transition described by the Gibbs free energy, $r_c$ must jump with $r_h$ at the phase transition, regardless of whether $r_c(r_h)$ is monotonic or not. This is because the latent heat is nonzero in a first-order phase transition, which forces the black hole mass to change and hence leads to a discontinuous change in $r_c$. It is worth noting that if a black hole has a second-order phase transition, then at the critical point the latent heat is zero, and $r_h$ and $r_c$ do not jump. This result not only fills the gap in previous studies for the nonmonotonic case of $r_c(r_h)$, but also further reveals the deeper reason behind the link between black hole phase transitions and particle circular orbits. For a zeroth-order phase transition, there is no simple universal conclusion. However, under mild assumptions one can still obtain a strong result: if the changes in the Gibbs free energy and in the entropy across the phase transition are not both zero and have the same sign, then $r_c$ must also jump with $r_h$. For these two types of phase transitions, the change of $r_c$ across the phase transition can be determined exactly from eq.~\eqref{eq:drc}. In our example, along isotherms of the RN--AdS black hole, the first-order phase transition indeed induces a jump in $r_c$, and the jump disappears at the critical point, as shown in figure~\ref{rpsiscot}.

The conclusion that $r_c$ must jump with $r_h$ at a phase transition provides a theoretical guarantee for potential observable effects of the transition. From a thermodynamic viewpoint, it also opens the possibility of using $\Delta r_c$ to replace $\Delta r_h$ as an order parameter. Although this possibility has been confirmed in many previous studies, a model-independent proof has been lacking. In this paper, we gave such a proof: near the thermodynamic critical point, the first law requires $r_c(r_h)$ to be locally monotonic, so the critical exponents remain unchanged under the reparametrization $r_h\to r_c$. This means that, in black hole thermodynamic evolution, processes characterized by $r_c$ and by $r_h$ belong to the same universality class. This result also suggests an intrinsic connection between black hole thermodynamics and spacetime geometry.

We would like to emphasize that, to our knowledge, the nonmonotonic behavior of $r_c(r_h)$ is reported here for the first time. If such nonmonotonicity is imprinted on the black hole shadow or accretion-disk images, we could in principle not only detect the occurrence of phase transitions but also qualitatively distinguish the specific thermodynamic evolution process that the black hole is undergoing. This finding would open up a new avenue for astronomical observations of black hole thermodynamics and warrants further investigation.

Before ending this paper, we would also like to highlight an intriguing observation. Our discussion of eqs.~\eqref{eq:mks1}--\eqref{eq:mks4} shows that replacing the independent variables in $P(V)$ and $T(S)$ with $r_h$ produces new equal-area relations [eqs.~\eqref{eq:mks3} and \eqref{eq:mks4}] that no longer guarantee that the $P(r_h)$ or $T(r_h)$ curve and the segment $AA'$ enclose two equal areas (see figure~\ref{prhtrh}). Even so, at a formal level, eqs.~\eqref{eq:mks3} and \eqref{eq:mks4} can still be regarded as ``Maxwell equal-area relations'' obtained from the respective reparameterizations $S \mapsto r_h$ and $V \mapsto r_h$. In other words, since both $V(r_h)$ and $S(r_h)$ are monotonic, we can not only qualitatively identify phase transition and critical point information from the $P(r_h)$ and $T(r_h)$ plots\footnote{The relations $P(r_h)$ and $T(r_h)$ can be derived from $P(V)$ and $T(S)$ through horizontal stretching, compression, or reflection (along the $V$- or $S$-axis), preserving their qualitative behavior (see Appendix~\ref{appa} for the mathematical details).}, but also quantitatively construct the line segment $AA'$ such that it satisfies the Maxwell equal-area law. This conclusion can be naturally extended to $P(r_c)$ and $T(r_c)$, provided that $r_c(r_h)$ is monotonic [which ensures that both $V(r_c)$ and $S(r_c)$ are also monotonic]. However, nonmonotonic $r_c(r_h)$ renders the analysis considerably more complicated. Deriving $P(r_c)$ and $T(r_c)$ requires inverting $r_c(r_h)$ to express $r_h$ as a function of $r_c$. Yet the nonmonotonicity implies that $r_h(r_c)$ is multivalued, consisting of at least two branches (such as a decreasing branch and an increasing branch). As a result, a single value of $r_c$ corresponds to at least two values of $r_h$, and thus typically to at least two distinct values of $P$ and two distinct values of $T$. Thus, strictly speaking, $P(r_c)$ and $T(r_c)$ are no longer ordinary single-valued functions and instead become multivalued curves parameterized by $r_h$. More explicitly, each value of $r_h$ maps to a point $\bigl(r_c(r_h),\,P(r_h)\bigr)$ in the $r_c$-$P$ plane and $\bigl(r_c(r_h),\,T(r_h)\bigr)$ in the $r_c$-$T$ plane. Geometrically, this transformation not only stretches, compresses, or reflectes the $P(V)$ and $T(S)$ curves but also introduces a kink at the points where $r_c(r_h)$ reaches its extrema. Consequently, the resulting curve remains continuous but loses differentiability at this point. In this case, the critical point of $P(V)$ and $T(S)$ is either preserved or becomes the point where a kink forms. At the same time, the $S$-shaped structure, or more generally the structure associated with the phase transition, may be destroyed. Therefore, from a qualitative perspective, these changes have minimal impact on one's ability to identify the critical point directly from the $P(r_c)$ and $T(r_c)$ plots, but they may affect one's ability to extract the phase transition information. From a quantitative perspective, the critical point will in most cases still satisfy the reparameterized critical point conditions. By contrast, for the Maxwell equal-area relations given by eqs.~\eqref{eq:mks1} and \eqref{eq:mks2}, the integral must be split into at least two segments, which means that a suitable line segment $AA'$ may not be found to satisfy eqs.~\eqref{eq:mks1} and \eqref{eq:mks2}. In this sense, the nonmonotonicity of $r_c(r_h)$ may obstruct any attempt to establish a formal Maxwell equal-area relation in terms of $r_c$ that parallels eqs.~\eqref{eq:mks3} and \eqref{eq:mks4}. Therefore, $r_c$ may not fully replace $r_h$ for phase transition analysis. The above conclusion can also be generalized to other quantities expressed as functions of $r_h$. For example, the authors of Ref.~\cite{Hale:2024lzh} reported a similar observation when replacing $r_h$ with the Lyapunov exponent. It should be emphasized that the above discussion of the Maxwell equal-area relation relies entirely on reparameterization, which is essentially a mathematical operation. Physically, however, the relation stems from the equality of thermodynamic potentials at the coexistence point of a phase transition. For example, integrating the differential relations $\dd G=-S \dd T=0$ and $\dd G=V \dd P=0$ directly yields the relations [eqs.~\eqref{eq:mks1} and \eqref{eq:mks2}]. Reparameterization-based relations [eqs.~\eqref{eq:mks3} and \eqref{eq:mks4}] are therefore just mathematical rewritings of these integrals, which is why we refer to them as formal rather than genuine ones. To derive a genuine Maxwell equal-area relation associated with $r_c$, one would need to identify a thermodynamic intensive quantity $R$ conjugate to $r_c$, such that the differential relation for some thermodynamic potential $\Psi$ takes the form
\begin{equation}
 0=\dd\Psi \propto r_c\, \dd R.
\end{equation}
However, $r_c$ itself is not a natural thermodynamic variable, and therefore we believe that it is not possible to construct such a conjugate quantity and the corresponding thermodynamic potential in a natural way. In fact, even for $r_h$ itself, one cannot naturally construct a corresponding conjugate quantity and thermodynamic potential. Overall, our view is that if $r_c(r_h)$ is nonmonotonic, then this may obstruct attempts to build a formal Maxwell equal-area relation based on $r_c$ for a quantitative analysis of phase transitions. At the same time, regardless of the monotonicity of $r_c(r_h)$, it is unlikely to establish a natural Maxwell equal-area relation associated with $r_c$. Nevertheless, these limitations do not prevent $r_c$ from qualitatively revealing phase transition information, since it must still exhibit a discontinuous jump alongside $r_h$ at the phase transition.

\acknowledgments

We would like to thank Zhongzhinan Dong and Jiawei Chen for helpful discussions. This work is supported in part by NSFC Grants No. 12165005 and No. 11961131013.

\appendix

\section{Preservation of functional behavior under monotonic reparameterization}
\label{appa}

Suppose that $F(x)$ is a differentiable function defined on an interval $I\subset\mathbb{R}$, and let $x_m$ denote its $m$-th root, so that $F(x_m)=0$. Now introduce a new variable $y$ and assume a smooth, strictly monotonic mapping
\begin{equation}
 x = x(y), \qquad y\in J\subset\mathbb{R},
\end{equation}
with ${\mathrm{d}x}/{\mathrm{d}y}\neq 0$ for all $y\in J$. We then reparameterize $F(x)$ as
\begin{equation}
 \widetilde{F}(y) = F\bigl(x(y)\bigr).
\end{equation}
Applying the chain rule gives
\begin{equation}
 \frac{\mathrm{d}\widetilde{F}}{\mathrm{d}y}
 = \frac{\mathrm{d}F}{\mathrm{d}x}\,\frac{\mathrm{d}x}{\mathrm{d}y}.
 \label{eq:a3}
\end{equation}
Since $\mathrm{d}x/\mathrm{d}y\neq0$, one immediately has
\begin{equation}
 \frac{\mathrm{d}\widetilde{F}}{\mathrm{d}y}=0
 \quad\Longleftrightarrow\quad
 \frac{\mathrm{d}F}{\mathrm{d}x}=0.
\end{equation}
Thus, the stationary points of $F$ and $\widetilde{F}$ are in one-to-one correspondence: each stationary point $x_m$ of $F$ maps to a stationary point $y_m=x^{-1}(x_m)$ of $\widetilde{F}$, and vice versa.

To demonstrate that the nature of the stationary points is preserved, we differentiate once more:
\begin{equation}
 \frac{\mathrm{d}^2\widetilde{F}}{\mathrm{d}y^2}
 =\frac{\mathrm{d}^2 F}{\mathrm{d}x^2}\left(\frac{\mathrm{d}x}{\mathrm{d}y}\right)^2
 +\frac{\mathrm{d}F}{\mathrm{d}x}\frac{\mathrm{d}^2 x}{\mathrm{d}y^2}.
\end{equation}
At a stationary point $y_m$ where ${\mathrm{d}F}/{\mathrm{d}x}=0$, the above equation reduces to
\begin{equation}
 \left.\frac{\mathrm{d}^2\widetilde{F}}{\mathrm{d}y^2}\right|_{y_m}
 = \left.\frac{\mathrm{d}^2 F}{\mathrm{d}x^2}\right|_{x_m}\left.\left(\frac{\mathrm{d}x}{\mathrm{d}y}\right)^2\right|_{y_m}.
\end{equation}
Since $\left({\mathrm{d}x}/{\mathrm{d}y}\right)^2>0$, it follows that
\begin{equation}
 \begin{aligned}
 \left.\frac{\mathrm{d}^2 F}{\mathrm{d}x^2}\right|_{x_m}>0 &\ \Longrightarrow\
 \left.\frac{\mathrm{d}^2\widetilde{F}}{\mathrm{d}y^2}\right|_{y_m}>0,\\[4pt]
 \left.\frac{\mathrm{d}^2 F}{\mathrm{d}x^2}\right|_{x_m}<0 &\ \Longrightarrow\
 \left.\frac{\mathrm{d}^2\widetilde{F}}{\mathrm{d}y^2}\right|_{y_m}<0.
 \end{aligned}
\end{equation}
Hence, under any smooth and monotonic reparameterization $x\mapsto y$, a local minimum (maximum) of $F$ remains a local minimum (maximum) of $\widetilde{F}$. Away from the stationary points, eq.~\eqref{eq:a3} implies that when $\mathrm{d}x/\mathrm{d}y>0$, the signs of $\mathrm{d}\widetilde{F}/\mathrm{d}y$ and $\mathrm{d}F/\mathrm{d}x$ coincide, and are opposite when $\mathrm{d}x/\mathrm{d}y<0$. Geometrically, the former corresponds to a horizontal stretching of the original curve, while the latter involves both stretching and a reflection about the $x$-axis. In either case, the overall qualitative profile of the function remains unchanged.

If $F$ is identified with the thermodynamic temperature $T$ or pressure $P$ of a black hole, and the variables are smoothly and monotonically related as $x=r_h$ and $y=r_c$, then under the mapping $r_h \mapsto r_c$ the functions $T(r_c)$ and $P(r_c)$ retain their characteristic van der Waals-like $S$-shaped structure. As a result, during the small to large black hole phase transition, the critical radius $r_c$ undergoes a finite discontinuous jump.

\section{Conditions of the photon sphere and ISCO}
\label{appg}

For photons ($\epsilon=0$) the radial function~\eqref{eq:rd} reduces to
\begin{equation}
 \mathcal{R}(r)=\frac{g(r)}{f(r)}E^{2}-\frac{g(r)}{r^{2}}L^{2}.
\end{equation}
Imposing the circular orbit conditions, $\mathcal{R}(r_{\rm ps})=\mathcal{R}'(r_{\rm ps})=0$, gives
\begin{align}
 \mathcal{R}(r_{\rm ps})&=\frac{g(r_{\rm ps})}{f(r_{\rm ps})}E^{2}-\frac{g(r_{\rm ps})}{r_{\rm ps}^{2}}L^{2}=0,\\
 \mathcal{R}'(r_{\rm ps})&=
 \left[\frac{g(r_{\rm ps})}{f(r_{\rm ps})}\right]'E^{2}
 -\frac{g'(r_{\rm ps})}{r_{\rm ps}^{2}}L^{2}
 +\frac{2g(r_{\rm ps})}{r_{\rm ps}^{3}}L^{2}=0.
\end{align}
Eliminating $E^{2}$ and $L^{2}$ yields
\begin{equation}
 \frac{\Phi_{0}(r_{\rm ps})}{r_{\rm ps}f(r_{\rm ps})}=0,
\end{equation}
where we define
\begin{equation}
 \Phi_{0}(r_{\rm ps}):=r_{\rm ps}\,f'(r_{\rm ps})-2f(r_{\rm ps}),
\end{equation}
which corresponds to the photon sphere equation~\eqref{eq:phi0}. Imposing the instability of the photon circular orbit, $\mathcal{R}''(r_{\rm ps})>0$, gives
\begin{equation}
 \mathcal{R}''(r_{\rm ps})
 =-\frac{f(r_{\rm ps})g(r_{\rm ps})L^2}{r_{\rm ps}}\,
 \underbrace{\big[r_{\rm ps}f''(r_{\rm ps})-f'(r_{\rm ps})\big]}_{\;\equiv\ \partial_r\Phi_0(r_{\rm ps})}>0.
\end{equation}
Since $f(r_{\rm ps})g(r_{\rm ps})L^2/r_{\rm ps}>0$, it follows that
\begin{equation}
 \Gamma_0
 :=\partial_r\Phi_0(r_{\rm ps})
 =r_{\rm ps}f''(r_{\rm ps})-f'(r_{\rm ps})<0,
\end{equation}
which corresponds to the stability criterion~\eqref{eq:g0}.

For massive particles ($\epsilon=1$) the radial function~\eqref{eq:rd} reduces to
\begin{equation}
 \mathcal{R}(r)=\frac{g(r)}{f(r)}E^{2}-g(r)\!\left(1+\frac{L^2}{r^2}\right).
\end{equation}
The conditions
\begin{equation}
 \mathcal{R}(r_{\rm isco})
 =\mathcal{R}'(r_{\rm isco})
 =\mathcal{R}''(r_{\rm isco})=0,
\end{equation}
give three algebraic equations for $E^{2}$, $L^{2}$, and $r_{\rm isco}$:
\begin{align}
 \mathcal{R}(r_{\rm isco})&=\frac{g(r_{\rm isco})}{f(r_{\rm isco})}E^{2}-g(r_{\rm isco})\!\left(1+\frac{L^2}{r_{\rm isco}^2}\right)=0,\\
 \label{eq:c11}
 \mathcal{R}'(r_{\rm isco})
 &= \left[\frac{g(r_{\rm ps})}{f(r_{\rm ps})}\right]'E^{2}
 +\frac{2g(r_{\rm isco})}{r_{\rm isco}^3}L^2
 -g'(r_{\rm isco})\!\left(1+\frac{L^2}{r_{\rm isco}^2}\right)=0,\\
 \mathcal{R}''(r_{\rm isco})
 &= \left[\frac{g(r_{\rm ps})}{f(r_{\rm ps})}\right]''E^{2}
 -\frac{6g(r_{\rm isco})L^2}{r_{\rm isco}^4}+\frac{4g'(r_{\rm isco})L^2}{r_{\rm isco}^3}
 -g''(r_{\rm isco})\!\left(1+\frac{L^2}{r_{\rm isco}^2}\right)=0.
\end{align}
Eliminating $E^{2}$ yields
\begin{equation}
 \mathcal{R}''(r_{\rm isco})=
 -\frac{2L^2\,\Phi_1(r_{\rm isco})}{r_{\rm isco}^4\,f(r_{\rm isco})\,f'(r_{\rm isco})}=0,
 \label{eq:hr}
\end{equation}
where we define
\begin{equation}
 \begin{aligned}
 \Phi_1(r_{\rm isco})
 := r_{\rm isco}\,g(r_{\rm isco})\,f(r_{\rm isco})\,f''(r_{\rm isco})
 &- 2\,r_{\rm isco}\,g'(r_{\rm isco})\,f(r_{\rm isco})\,f'(r_{\rm isco})
 + 3\,g(r_{\rm isco})\,f(r_{\rm isco})\,f'(r_{\rm isco})\\
 &+2\,r_{\rm isco}\,g'(r_{\rm isco})\,f(r_{\rm isco})
 -2\,r_{\rm isco}\,g(r_{\rm isco})\,f'(r_{\rm isco})
 \end{aligned}
\end{equation}
which corresponds to the ISCO equation~\eqref{eq:phi1}. To determine the sign of
\begin{equation}
 \Gamma_1
 :=\partial_r\Phi_1(r_{\rm isco}),
\end{equation}
we rewrite eq.~\eqref{eq:hr} as
\begin{equation}
 \mathcal{R}''(r_{\rm c})=
 -\frac{2L^2\,\Phi_1(r_{\rm c})}{r_{\rm c}^4\,f(r_{\rm c})\,f'(r_{\rm c})},
\end{equation}
where $r_{\rm c}$ denotes the radius of a circular orbit in the neighborhood of the ISCO. By definition of the ISCO, orbits with $r_{\rm c}>r_{\rm isco}$ are stable $\bigl[\mathcal{R}''(r_{\rm c})<0\bigr]$, whereas those with $r_{\rm c}<r_{\rm isco}$ are unstable $\bigl[\mathcal{R}''(r_{\rm c})>0\bigr]$. Next, we analyze the sign of $\mathcal{R}''(r_{\rm c})$. Replacing $r_{\rm isco}$ with $r_{\rm c}$ in eq.~(\ref{eq:c11}) and simplifying, one obtains
\begin{equation}
 f(r_{\rm c})\,f'(r_{\rm c})
 =\frac{2\,f(r_{\rm c})^2\,L^2}{r_{\rm c}\bigl(r_{\rm c}^2+L^2\bigr)}
 >0,
\end{equation}
which implies $-2L^2\bigl[r_{\rm c}^4 f(r_{\rm c}) f'(r_{\rm c})\bigr]^{-1}<0$. Therefore,
\begin{equation}
 \mathrm{sgn}\!\left[\mathcal{R}''(r_{\rm c})\right]
 =-\,\mathrm{sgn}\!\left[\Phi_1(r_{\rm c})\right].
\end{equation}
Expanding $\Phi_1$ about the ISCO,
\begin{equation}
 \Phi_1(r_{\rm c})=\Phi_1(r_{\rm isco})
 +\partial_r\Phi_1(r_{\rm isco})\,(r_{\rm c}-r_{\rm isco})
 +O\bigl[(r_{\rm c}-r_{\rm isco})^2\bigr],
\end{equation}
and using $\Phi_1(r_{\rm isco})=0$, we obtain near $r_{\rm isco}$:
\begin{equation}
 \mathrm{sgn}\left[\mathcal{R}''(r_{\rm c})\right]
 =-\mathrm{sgn}\!\left[\partial_r\Phi_1(r_{\rm isco})\,(r_{\rm c}-r_{\rm isco})\right].
\end{equation}
Consequently, to ensure $\mathcal{R}''(r_{\rm c})>0$ for $r_{\rm c}<r_{\rm isco}$ and $\mathcal{R}''(r_{\rm c})<0$ for $r_{\rm c}>r_{\rm isco}$, the radial derivative of $\Phi_1$ at the ISCO must be positive, i.e., $\Gamma_1=\partial_r\Phi_1(r_{\rm isco})>0$, which is precisely the stability criterion~\eqref{eq:g1}.

\section{Proof of the positivity of $\partial_P\Phi_1$}
\label{appd}

The full expression of $\partial_P\Phi_1$ reads
\begin{equation}
 \begin{aligned}
 \partial_P\Phi_1
 = \frac{16\pi\, r_{\mathrm{isco}}}{(d-2)(d-1)}
 \Bigl[\,8 + 4(d-1)(d-2)q^{2}r_{\mathrm{isco}}^{6-2d}
 - c_{d}(d-1)(d+1)M\,r_{\mathrm{isco}}^{3-d}\Bigr].
 \end{aligned}
\end{equation}
Since the prefactor is positive, the sign of $\partial_P\Phi_1$ is determined by the bracketed term. We define $x:=r_{\mathrm{isco}}^{\,d-3}>0$, so that the bracketed term can be written in terms of $x$ as
\begin{equation}
 8+\frac{4(d-1)(d-2)q^2}{x^2}-\frac{(d-1)(d+1)c_{d} M}{x}.
\end{equation}
To simplify the analysis, we multiply the above expression by $x^2>0$, which does not change the sign of the bracketed term. We then obtain an upward-opening quadratic function,
\begin{equation}
 \mathcal{Q}(x)
 := 8x^2 - (d-1)(d+1)c_d M x
 + 4(d-1)(d-2)q^2,
 \label{eq:d3}
\end{equation}
with discriminant
\begin{equation}
 \Delta_x
 = \big[(d-1)(d+1)c_d M\big]^2
 - 128(d-1)(d-2)q^2.
 \label{eq:d4}
\end{equation}
Therefore, if $\Delta_x<0$, then $\mathrm{sgn}\!\left(\partial_P\Phi_1\right)=\mathrm{sgn}(\mathcal Q(x))$ is positive for an arbitrary $x$. If instead $\Delta_x>0$, then
\begin{equation}
 \mathrm{sgn}\left(\partial_P\Phi_1\right) =\mathrm{sgn}(\mathcal Q(x)) =
 \begin{cases}
 +,& x\in(0,x_-)\cup(x_+,+\infty),\\[2pt]
 0,& x=x_\pm,\\[2pt]
 -,& x\in(x_-,x_+).
 \end{cases}\label{eq:sgnpartialPhi}
\end{equation}
Here
\begin{equation}
 x_\pm
 = \frac{(d-1)(d+1)c_d M
 \pm \sqrt{\Delta_x}}{16}
 \label{eq:d5}
\end{equation}
denote the two real roots of $\mathcal Q(x)$. To determine the sign of $\partial_P\Phi_1$, we first examine the sign of $\Delta_x$. Under the isothermal constraint one has
\begin{equation}
 T_0=\frac{(d-1)c_d M\,r_h^{d+3}-2(d-2)q^2 r_h^{6}-2 r_h^{2d}}{4\pi r_h^{1+2d}}.
\end{equation}
The requirement $T_0>0$ is equivalent to
\begin{equation}
 (d-1)c_d M\,r_h^{d+3}-2(d-2)q^2 r_h^{6}-2 r_h^{2d}>0.
\end{equation}
Introducing $s:=r_h^{d-3}$, we can rewrite the above expression in terms of $s$ as
\begin{equation}
 \mathcal{B}(s):=-2s^2+(d-1)c_d\,M\,s-2(d-2)q^2>0,
\end{equation}
which is a downward-opening quadratic function. For $\mathcal{B}(s)>0$ to hold, its discriminant must satisfy
\begin{equation}
 \Delta_s=\left[c_d(d-1)M\right]^2-16(d-2)q^2>0,
\end{equation}
implying
\begin{equation}
 M>\frac{4\sqrt{d-2}}{c_d(d-1)}\,q.
\end{equation}
Substituting this inequality into eq.~\eqref{eq:d4} gives
\begin{equation}
 \Delta_x>16\left(d-2\right)\left(d-3\right)^2\,q^2>0.
\end{equation}
Hence, $\partial_P\Phi_1$ may take positive, negative, or zero values shown in eq. \eqref{eq:sgnpartialPhi}, depending on the relative magnitude of $x$ and $x_\pm$. To clarify the relation between $x$ and $x_\pm$, we rewrite the expression of $\Phi_1(r;\lambda)$ in eq. \eqref{eq:phi1} with $\lambda=(M,q,P,d)$ in $d$-dimensional AdS black hole spacetime as
\begin{equation}
 \Phi_1(r;M,q,P,d)
 = P\,\partial_P\Phi_1 + \widetilde\Phi_1(r;M,q,d).
 \label{eq:d13}
\end{equation}
Fixing $M$, $q$, and $d$ at finite values and taking the limit $P\to +\infty$, the condition $\Phi_1(r;M,q,P,d)=0$ can be satisfied only if
\begin{equation}
 \partial_P\Phi_1 \to 0.
\end{equation}
This shows that the quantities $x_\pm^{1/(d-3)}$ correspond exactly to the limiting values of $r_{\rm isco}$ as $P\to +\infty$. Next, we examine the limit $P \to 0$. For $d=4$, one obtains
\begin{equation}
 \Phi_1=\frac{2M r_{\mathrm{isco}}^{2}(r_{\mathrm{isco}}-6 M) + 18 M q^2 r_{\mathrm{isco}}-8 q^4}{r_{\mathrm{isco}}^{5}}.
\end{equation}
The analysis can be divided into two cases:
\begin{itemize}
\item $q=0$: For the Schwarzschild black hole, this yields $r_{\mathrm{isco}}=6M$. In this case, the two roots are $x_\pm={15M}/{4}$ and $0$, and clearly $x\!\mid_{P=0}=6M>x_+$.
\item $q\neq0$: For the RN black hole, it is well known that $r_{\mathrm{isco}}$ attains its minimum at $M=q$, where $r_{\mathrm{isco,min}}=4M$. From eq.~\eqref{eq:d5}, $x_+$ decreases monotonically with $q$, reaching its maximum $x_{+,{\rm max}}={15M}/{4}$ at $q=0$, so that $x_{\rm min}\!\mid_{P=0}=4M>x_{+,{\rm max}}={15M}/{4}$.
\end{itemize}
Therefore, for both the Schwarzschild and RN black holes one always has $x\!\mid_{P=0}>x_+$, and consequently $\partial_P\Phi_1>0$ in this limit.

Next, consider increasing $P$ from zero in eq.~\eqref{eq:d13}. Initially, one has $\Phi_1=\widetilde\Phi_1(r_{\mathrm{isco}}\!\mid_{P=0};M,q,d)=0$ with $\Gamma_1=\partial_r\widetilde\Phi_1>0$. Once $P$ increases, the term $\partial_P\Phi_1>0$ appears, so $\widetilde\Phi_1$ must decrease to maintain $\Phi_1=0$. Since $\Gamma_1>0$, this requires $r_{\mathrm{isco}}$ to decrease, meaning that $x$ moves toward $x_+$. As $P$ continues to increase, we assume continuous dependence of $r_{\mathrm{isco}}$ and $\partial_P\Phi_1$ on $P$, so that $x$ must vary continuously from $x\!\mid_{P=0}$ to $x_+\!\mid_{P\to +\infty}$. Consequently, for any finite $P$, the corresponding $x$ must lie within the interval $(x_+,\,x\!\mid_{P=0})$, which guarantees $\partial_P\Phi_1>0$ throughout. From a physical viewpoint, this reflects the suppressive effect of pressure on the ISCO radius: as $P$ increases, $r_{\mathrm{isco}}$ decreases monotonically toward $x_+$.

For higher dimensions ($d>4$), the argument becomes even simpler. At $P=0$, Schwarzschild and RN black holes possess no ISCO ($r_{\mathrm{isco}}\to +\infty$). As $P$ increases, $r_{\mathrm{isco}}$ moves inward from infinity toward $x_+$, again ensuring $\partial_P\Phi_1>0$. Hence, in all cases we conclude that
\begin{equation}
 \partial_P\Phi_1>0.
\end{equation}
The limit used above,
\begin{align}\label{eq:riso-limit}
 \lim_{P\to0} r_{\rm isco}=+\infty,
\end{align}
can be argued as follows. For the higher-dimensional Schwarzschild black hole ($q=0$, $d>4$), the ISCO condition takes the form
\begin{equation}
 \begin{aligned}
 \Phi_1
 =
 \frac{16\pi P\,r_{\rm isco}}{(d-2)(d-1)}
 \Bigl[\,8 - c_d(d^{2}-1)M\,r_{\rm isco}^{3-d}\Bigr]
 - c_d(d-3)M\,r_{\rm isco}^{2(1-d)}
 \Bigl[\,c_d(d-1)M\,r_{\rm isco}^{3}
 +(d-5)\,r_{\rm isco}^{d}\Bigr].
 \end{aligned}
\end{equation}
If $r_{\rm isco}$ remains finite as $P\to0$, the first term (pressure-dependent) vanishes and $\Phi_1$ reduces to
\begin{equation}
 \begin{aligned}
 \Phi_1
 =- c_d(d-3)M\,r_{\rm isco}^{2(1-d)}
 \Bigl[\,c_d(d-1)M\,r_{\rm isco}^{3}
 +(d-5)\,r_{\rm isco}^{d}\Bigr],
 \end{aligned}
\end{equation}
which admits no positive root for $r_{\rm isco}$ when $d>4$. Thus a finite $r_{\rm isco}$ is incompatible with the ISCO condition in this limit. By contrast, if $r_{\rm isco}\to +\infty$ as $P\to0$, the second term in $\Phi_1$ is suppressed, and one finds
\begin{equation}
 \Phi_1
 \;\simeq\;
 \frac{128\pi P\,r_{\rm isco}}{(d-2)(d-1)},
\end{equation}
so that the product $P\,r_{\rm isco}$ can tend to zero as $P\to0$. Therefore the ISCO condition can be satisfied only if
\begin{equation}
 \lim_{P\to0} r_{\rm isco}=+\infty.
\end{equation}
For the higher-dimensional RN black hole ($q\neq 0$, $d>4$), the ISCO condition takes the form
\begin{equation}
 \begin{aligned}
 \Phi_1
 =&-2(d-3)\,q\,r_{\rm isco}^{\,2(1-2d)}
 \bigl(-q\,r_{\rm isco}^3 + r_{\rm isco}^d\bigr)^2
 \bigl[2(d-2)\,q\,r_{\rm isco}^3 + (d-5)\,r_{\rm isco}^d\bigr] \\
 &+ \frac{32\pi P\,r_{\rm isco}}{(d-2)(d-1)}
 \Bigl\{4 - (d-1)\,q\,r_{\rm isco}^{\,3-2d}
 \bigl[-2(d-2)\,q\,r_{\rm isco}^3
 + (d+1)\,r_{\rm isco}^d\bigr]\Bigr\},
 \end{aligned}
\end{equation}
where we have imposed the extremality condition $M=2q/c_d$. Repeating the same analysis as in the Schwarzschild case then yields eq.~\eqref{eq:riso-limit}.

\section{Asymptotic relations in the small-horizon limit}
\label{appe}

We now examine the asymptotic behavior of $V\,\partial_M\Phi_1$ and $\partial_P\Phi_1$ in the small-horizon limit ($r_h\to0^+$), focusing on their leading divergent terms.

For $\partial_P\Phi_1$, since $r_h \to 0^+$ implies $P \to +\infty$, one has $\partial_P\Phi_1 \to 0$ (see Appendix~\ref{appd} for details). The thermodynamic volume takes the simple form
\begin{equation}
 V = \frac{\omega_{d-2}\, r_h^{\,d-1}}{d-1},
\end{equation}
and therefore vanishes as $V \sim r_h^{\,d-1} \!\to\! 0$. $\partial_M\Phi_1$ reads
\begin{equation}
 \begin{aligned}
 \partial_M\Phi_1=3c_d(d-1)(d-3)q^2 r_{\mathrm{isco}}^{8-3d}
 &-2c_d^2(d-1)(d-3)M\,r_{\mathrm{isco}}^{5-2d}\\
 &-c_d(d-3)(d-5)r_{\mathrm{isco}}^{2-d}
 -\frac{16c_d(d+1)\pi P\,r_{\mathrm{isco}}^{4-d}}{d-2},
 \label{eq:e2}
 \end{aligned}
\end{equation}
and its asymptotic behavior can be inferred once the leading scalings of $M$, $P$, and $r_{\mathrm{isco}}$ are determined. Under the isothermal constraint, one finds
\begin{equation}
 M(r_h)
 =\frac{\bigl(1 + 2\pi T_0 r_h\bigr)\,r_h^{2(d-3)} + 2(d-2)\,q^{2}}
 {c_d\,(d-1)\,r_h^{\,d-3}},
\end{equation}
and
\begin{equation}
 P(r_h)
 =\frac{(d-2)\!\left[\bigl(3-d + 4\pi T_0 r_h\bigr)\,r_h^{2(d-3)}+(d-3)\,q^{2}\right]}
 {16\pi\,r_h^{\,2(d-2)}}.
\end{equation}
In the small-horizon limit ($r_h\!\to\!0^+$), these behave as
\begin{equation}
 M \sim \frac{1}{r_h^{\,d-3}},
 \qquad
 P \sim \frac{1}{r_h^{\,2(d-2)}}.
 \label{eq:e5}
\end{equation}
For the ISCO, the divergence $P\!\to\!+\infty$ implies (see Appendix~\ref{appd})
\begin{equation}
 r_{\mathrm{isco}}^{\,d-3}\to x_+
 =\frac{(d-1)(d+1)c_{d} M + \sqrt{\Delta_x}}{16},
\end{equation}
where
\begin{equation}
 \Delta_x=\big[(d-1)(d+1)c_{d} M\big]^2
 -128(d-1)(d-2)q^2.
\end{equation}
Near $r_h\!\to\!0^+$, the dominant contribution arises from the mass term $M$, so that
\begin{equation}
 r_{\mathrm{isco}} \sim M^{\tfrac{1}{d-3}} \sim \frac{1}{r_h}.
 \label{eq:e8}
\end{equation}
Substituting eqs.~\eqref{eq:e5} and \eqref{eq:e8} into eq.~\eqref{eq:e2} gives $\partial_M\Phi_1 \sim -r_h^{-d}$, and hence
\begin{equation}
 V\,\partial_M\Phi_1 \sim -\frac{1}{r_h} \;\longrightarrow\; -\infty.
\end{equation}



\end{document}